\documentclass[english]{article}
\usepackage{arxiv}
\usepackage{graphicx}
\usepackage[T1]{fontenc}
\usepackage{geometry}
\usepackage{amsmath}
\usepackage{babel}
\usepackage{color}
\usepackage[normalem]{ulem}
\usepackage{amssymb}
\usepackage{hyperref}
\usepackage[title]{appendix}
\usepackage{authblk}
\usepackage{siunitx} 
\usepackage{booktabs}

\begin{document}
\title{\textbf{Direct numerical simulation of a moist cough flow using Eulerian approximation for liquid droplets}}

\author[1]{\textbf{Rohit Singhal}}
\author[2]{\textbf{S. Ravichandran}}
\author[1,\thanks{Email address for correspondence : \texttt{sdiwan@iisc.ac.in}}]{\textbf{Sourabh S. Diwan}}
\affil[1]{Department of Aerospace Engineering, Indian Institute of Science, Bengaluru 560012, India}

\affil[2]{Nordic Institute for Theoretical Physics, KTH Royal Institute of Technology and Stockholm University, Stockholm, SE-106 91, Sweden}


\maketitle
\begin{abstract}
The COVID-19 pandemic has inspired several studies on the fluid dynamics of respiratory events. Here, we propose a computational approach in which respiratory droplets are coarse-grained into an Eulerian liquid field advected by the fluid streamlines. A direct numerical simulation is carried out for a moist cough using a closure model for space-time dependence of the evaporation time scale. Estimates of the Stokes number are provided, for the initial droplet size of $10\mu$m, which are found to be $<<1$ thereby justifying the neglect of droplet inertia. Several of the important features of the moist-cough flow reported in the literature using Lagrangian tracking methods have been accurately captured using our scheme. Some new results are presented, including the evaporation time for a ``mild'' cough, a saturation-temperature diagram and a favourable correlation between the vorticity and liquid fields. The present approach is particularly useful for studying the long-range transmission of virus-laden droplets. 
\end{abstract}


\section{Introduction}
The airborne transmission of respiratory infection is thought to be responsible for the COVID-19 pandemic that is afflicting the world. The SARS-CoV-2 virus responsible for the contagion spreads through air, not only by people with respiratory symptoms (coughing/sneezing)  \cite{Bourouiba2020}, but also by asymptomatic carriers through speech and even breathing \cite{morawska2020airborne}.
As a result, the fluid and droplet dynamics of the various respiratory events has been a subject of several investigations, especially since the start of the pandemic. These include studies on symptomatic respiratory events such as coughs or sneezes \cite{Lohse2021,Rosti2020,fabregat2021direct, domino2021case}, as well as those on everyday activities like breathing, talking, or singing  \cite{morawska2009size,Chao2009,Abkarian2020, singhal2021virus}.

Airborne transmission occurs through the transport of tiny, virus-laden liquid droplets expelled during respiratory events \cite{Bourouiba2021,mittal2020flow}.
These droplets vary in size from a few micrometers to hundreds of micrometers \cite{Duguid1946,somsen2020small}. The large droplets ($\mathcal{O}(100-1000)\mu$m) settle rapidly and are thus removed from the flow over distances $\mathcal{O}(1\mathrm{m})$ \cite{Bourouiba2014}. (They may survive on the surfaces where they land, leading to ``fomite'' transmission; this was believed, at the beginning of the pandemic, to be an important mode of transmission of SARS-CoV-2  \cite{Asadi2020,Morawska2020}.) On the other hand, since small droplets remain suspended in the respiratory flow for longer times, they are transported over larger distances. The turbulence present in respiratory flows (which can be effectively modelled as turbulent puffs \cite{Bourouiba2014}) plays an important role in determining the trajectories of such droplets. Sufficiently small droplets (called aerosols) may, in fact, remain suspended and be transmitted through ventilation systems in large buildings \cite{Qian2010,zhao2005numerical,chen2010some}. Since the early work of Wells \cite{WELLS1934}, human respiratory flows have been studied experimentally to determine the droplet size distribution and typical flow velocities at the mouth, and the subsequent flow and droplet evolution; for more recent studies, see Refs. \cite{Abkarian2020,Bourouiba2014,gupta2009flow,Stadnytskyi2020}. The experimental studies have also served to provide useful inputs to numerical investigations, e.g., for determining the parameters of simulation.

Numerical studies of the dynamics of droplets over distances larger than a few metres are computationally expensive and typically utilise the Reynolds-averaged Navier-Stokes (RANS) equations which involve ad-hoc models for the turbulent viscosities and diffusivities; a typical example is the building ventilation, which involves large length and time scales ($\mathcal{O}(10\mathrm{m})$ and $\mathcal{O}(10^3 \mathrm{s})$ \cite{Qian2010,Shao2021}. Similar approaches have also been used in studying the effects of nonzero ambient flow (such as presence of wind) on the dispersion of virus-laden droplets in individual respiratory events \cite{dbouk2020coughing,feng2020influence}. Note that the individual respiratory flows that travel up to a distance of 2m are amenable to more accurate simulation methods such the large-eddy simulation (LES) and the direct numerical simulation (DNS). The former resolves the large energy-carrying scales in the flow while parameterising the smaller scales \cite{Abkarian2020, balachandar2021investigation}, whereas the latter aims to resolve scales sufficiently small (ideally up to the Kolmogorov scale) so that the dissipation of energy is captured accurately \cite{moin1998direct}; see Refs. \cite{Lohse2021, Rosti2020}. Such detailed numerical studies, combined with experimental studies, offer better insight into the survival and transmission of droplets, which may in turn help improve the respiratory-flow models in the large-scale studies, such as ventilation.

Human respiratory flows are essentially multi-phase flows in which the evaporation and condensation processes  play an important role in determining the droplet lifetime \cite{ Lohse2021, WELLS1934}. Such processes can also affect the buoyancy of the respiratory flow through the release/absorption of the latent heat. It is therefore important to incorporate the thermodynamics of phase change in a numerical simulation of such flows. However, for flows generated during human speech, the droplets are typically very small ($\mathcal{O}(5\mu$m)), and evaporate fast to convert into droplet nuclei in the form of dissolved solid substances which can continue to harbour virions. Such flows have been simulated without incorporating phase changes and using a scalar field to mimic the transport of the droplet nuclei \cite{Abkarian2020, singhal2021virus}. 

On the other hand, in flows generated by violent respiratory events like coughs or sneezes, a wide range of droplet sizes is present and there is a considerable variation in the maximum droplet size expelled during a cough or a sneeze from one person to another \cite{bourouiba2021review}. As a result, for such flows, the phase changes of liquid droplets into a vapour and vice-a-versa are critical for understanding the flow evolution and droplet dynamics; another relevant aspect is the gravitational settling of large droplets, typically greater than $100 \mu$m in diameter. Recent numerical studies have investigated these aspects, including the role of turbulence on thermodynamics and droplet motion. Chong et al. \cite{Lohse2021} performed DNS of a turbulent vapour puff and showed that the relative humidity and temperature of the ambient affect the longevity of the liquid water droplets; see also Ref. \cite{LohsePRF2021growth}. They carried out simulations for 50\% and 90\% ambient relative humidity and found presence of supersaturation in the flow for the latter case, which promoted an initial droplet growth resulting in extended droplet lifetimes. Rosti et al. \cite{Rosti2020} compared the results on the evolution of droplets obtained from a DNS with those obtained from a coarse-DNS (i.e., after filtering out  small-scale fluctuations)  and showed that the latter can considerably underestimate the lifetime of droplets. Liu et al. \cite{balachandar2021investigation}, in their LES study, found that a portion of their simulated moist puff separated from the main flow and travelled along a random direction at a faster speed. They also proposed a theory for predicting puff size, velocity, distance travelled and droplet size distribution, and compared the predictions with their LES results.

The above DNS and LES studies have employed a coupled Eulerian-Lagrangian approach to solve for the fluid and droplet velocities respectively; see also Ref.  \cite{fabregat2021direct}. These simulations have provided a wealth of useful information and have helped enhance our understanding of the dynamics of respiratory flows. However, since this approach involves tracking each individual droplet, the data that needs to be handled can become exceedingly large, especially when the respiratory event expels a large number of droplets of moderate size (which can be as large as $10^5$ droplets per $\mathrm{cm}^3$ \cite{bourouiba2021review}). This also makes such simulations computationally expensive, inherently limiting their use in numerical experiments and parametric studies. In this work, we use a middle ground between the  Lagrangian particle tracking methods, and RANS simulations, by coarse-graining the liquid water droplets into an Eulerian field that is carried with the flow, while resolving sufficiently small scales in the flow. This approach, which, in principle, solves for only one of the moments of the droplet distribution--the total liquid content--in the flow, is familiar in the atmospheric sciences as the ``one-moment'' scheme (e.g. \cite{AtmosLin1983bulk,Atmosgrabowski1998toward}; see Ref. \cite{beck2002development} which discusses the relevance of this method for analysing dense liquid sprays). Here we propose an ``extended one-moment'' scheme in which we interpret the liquid field in terms of a collection of droplets with the number density considered uniform in space (but varying in time), and the droplet radius a function of space and time. Since we resolve the small scales in the flow, the computational requirements remain much greater than that for RANS simulations of the same problem. However, our approach is algorithmically simpler than Lagrangian particle tracking while still solving the Navier-Stokes equations without approximation. 

We use this approach to study the typical flow produced by a ``mild'' cough, i.e., involving relatively low cough flow rates as in a ``throat-clearing'' cough, for example. The Navier-Stokes equation is solved within the Boussinesq approximation and the Clausis-Clayperon relation is used for relating local relative humidity to local temperature. The extended one-moment scheme provides a closure for the evaporation/condensation rates in the flow. The liquid content at the orifice (which mimics mouth opening) is prescribed to be consisting of mono-disperse droplets of $10\mu$m diameter, which are small enough to neglect droplet inertia. This assumption is supported \textit{a posteriori} by providing estimates of the Stokes number which are shown to be much smaller than unity. We carry out a careful comparison of our results with those from the literature obtained using Lagrangian particle tracking, and show that we are able to reproduce many of the important features of a moist cough flow, including the initial supersaturation. We also present some new results on the relative rates of decay of the saturation and temperature fields away from the orifice, and the interplay between the liquid content and vorticity fields.

The remainder of the paper is organised as follows. We describe the geometry of the problem and the governing evolution equations in section \ref{sec:numerical_detials}, including the treatment for incorporating the thermodynamics of phase change. Therein, we also provide numerical details for the present simulation. Section \ref{sec:results} reports simulation results wherein we first compare the results from the extended one-moment scheme with a more rudimentary model of treating evaporation time scale as a constant. This is followed by a detailed analysis of the data \textit{vis-a-vis} available results. We end section \ref{sec:results} by presenting some thoughts on the advantages/limitations of the proposed approach in the context of respiratory flows. Finally, the conclusions are presented in section \ref{sec:conclusion}.

\section{Governing equations and numerical details \label{sec:numerical_detials}}

\subsection{Geometry and problem setup \label{subsec:geometry}}

\begin{figure}[!h]
\centering%
\begin{minipage}{0.47\textwidth}
  \includegraphics[width=1.0\textwidth]{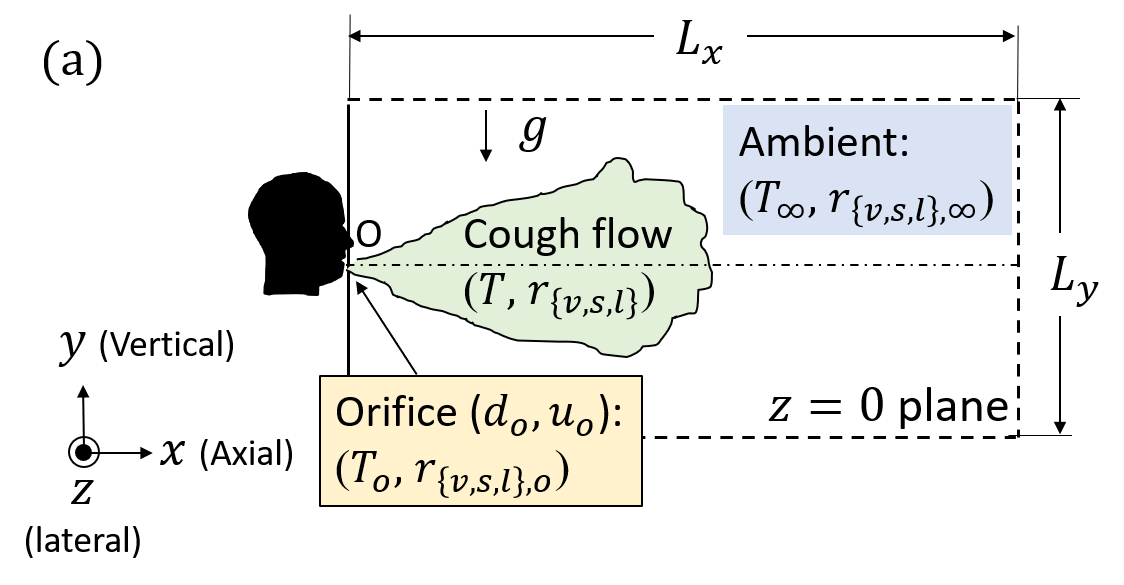}
\end{minipage}%
\hspace{0.03\textwidth}%
\begin{minipage}{0.48\textwidth}
  \includegraphics[width=1.0\textwidth]{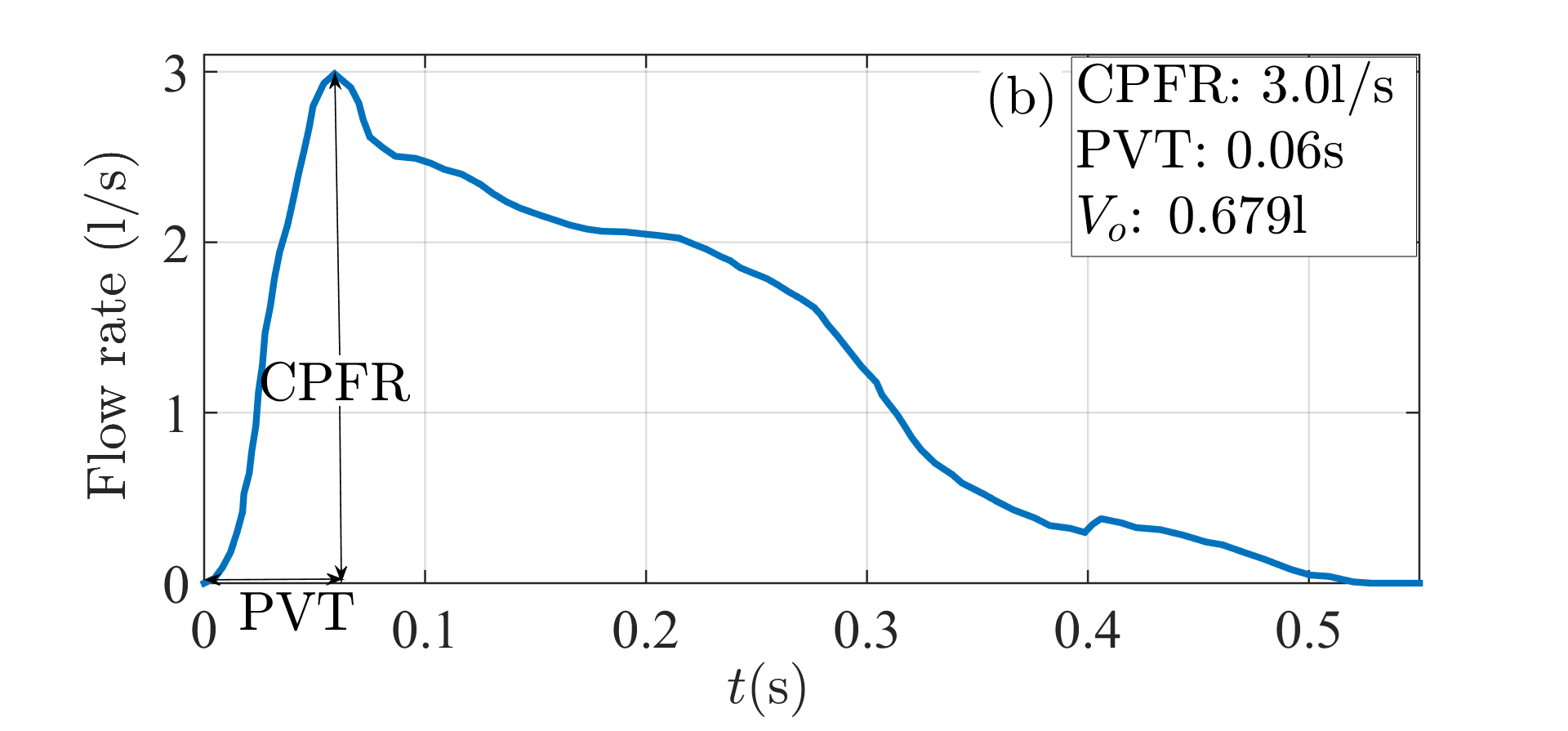}
\end{minipage}
\caption{(a) A Schematic side view ($z=0$ ) of the computational domain used to simulate a mild cough. A jet of mean velocity $u_0(t)$ exits the mouth, modelled as an orifice of diameter $d_0$.  
The temperature $T$ and the mixing ratios of the vapour and liquid $r_{v,l}$ at the orifice (subscript $_0$) and the ambient (subscript $_\infty$) are specified.
(b) The prescribed flow rate at the orifice used to calculate the inlet velocity $u_o$ at the orifice.}
\label{fig1}
\end{figure}

The domain used in the present numerical simulations, shown schematically in Figure \ref{fig1}a, is a cuboidal volume of dimensions $L_x\times L_y\times L_z$.  Here $x$, $y$ and $z$ are the axial, vertical (against gravity) and lateral co-ordinates respectively. Following Gupta et al. \cite{gupta2009flow}, the mouth of the person coughing is modelled as a circular orifice with a diameter $d_o=2.25$cm in the vertical $y-z$ plane, and is centred at the origin of coordinates (figure \ref{fig1}a). The inlet velocity $u_0$ at the orifice is obtained from the laboratory measurements of the flow rates in cough flows reported in Ref.\cite{gupta2009flow} (see  figure 5 in Ref. \cite{gupta2009flow} ).

Two quantities specify the flow-rate profile as a function of time at the orifice-- (a) the maximum flow rate or the cough peak flow rate (CPFR) and (b) the time of its occurrence or the peak-velocity time (PVT). A wide range of values have been reported for CPFR ($1.6-8.5$ litres/s) and PVT ($0.057-0.11$s); see \cite{gupta2009flow}. Here we choose CPFR$=3.0$ litres/s and PVT$=0.06$s representative of a ``mild'' cough, and the resulting flow rate as a function of time is shown in Figure~\ref{fig1}b. The total volume of fluid expelled from the orifice, the cough expiratory volume ($V_o$), is $0.679$ litres in a total cough time of $0.528$s; $V_o$ is the integral under the curve in figure \ref{fig1}b. The average velocity based on the orifice flow rate and total cough time is taken as the characteristic velocity scale, $u_c=3.232$ m/s.

Since we consider the thermodynamics of evaporation of the liquid droplets  expelled during a cough, the temperature and other thermodynamic variables at the orifice and in the ambient need to be specified. We label quantities at the orifice with a subscript $_o$ and ambient quantities with a subscript $_\infty$ (see figure \ref{fig1}a). We set the orifice temperature $T_o = 34^\circ $C to a value slightly lower than the typical body temperature and set the ambient temperature to a typical indoor temperature of $T_{\infty} = 20^\circ $C. These conditions are the same as those used by Chong et al.\cite{Lohse2021}; see also \cite{Bourouiba2014}. We use $T_\infty$ and $\Delta T_o = T_o-T_{\infty}$ as the characteristic scales respectively for temperature and temperature difference. The survival time of exhaled droplets increases as the temperature difference between the exhaled cough fluid and ambient indoor air, typically positive, increases \cite{Lohse2021}, with other parameters kept constant.

In addition to liquid droplets of a range of sizes, the exhaled fluid in a cough also contains water vapour, both mixed with dry air. The amounts of water substance in a cough is typically a small fraction by mass of the (dry) air ($10^{-5}-10^{-7}$;  \cite{Bourouiba2014}). We define the mixing ratios (mass per unit mass of dry air) of vapour as $r_v = \rho_v / \rho_d$ and liquid as $r_l = \rho_l / \rho_d$, where $\rho_d$, $\rho_v$ and $\rho_l$ are the densities of dry air, vapour and liquid water respectively. (We do not model the viscoelastic nature of the saliva droplets or the dissolved solid substances in these droplets.) The values of $r_v$ are more conveniently expressed in terms of the relative humidity $s=r_v/r_s$, where $r_s$ is the saturation vapour mixing ratio. For the present simulations, the humidity conditions are assigned as $s_\infty=0.9$ for the ambient, and $s_o=1$ at the orifice. These values are chosen so as to replicate one of the cases in Chong et al. \cite{Lohse2021}, who have used $s$ of 0.5 and 0.9 for the ambient and 1 for the orifice. This enables comparison of our results (especially the condensational growth due to supersaturation) obtained within the Eulerian field approximation with those in Ref. \cite{Lohse2021} who have done Lagrangian tracking of individual liquid droplets. Furthermore, $r_v$ and $r_l$ are also normalised using the saturation mixing ratio of the ambient, $r_{s,\infty}$  as the characteristic scale. We next describe the assumptions underlying the coarse-graining of the liquid water droplets into an Eulerian field.

\subsection{Eulerian treatment of droplets \label{subsec:tracer}}

Human coughs produce droplets of a wide range of sizes, a commonly-used distribution is the one provided by Duguid \cite{Duguid1946}, which ranges from ($2\mu$m-$1$mm); see  \cite{Lohse2021,Rosti2020,Bourouiba2014}. However, it is important to realize that no two coughs are identical, and  many different droplet-size distributions obtained from laboratory experiments on human subjects have been reported in the literature. Bourouiba \cite{bourouiba2021review} has compiled the available data which shows that coughs can have droplet concentrations as low as $0.1 \mathrm{cm^{-3}}$ and as high as $10^5 \mathrm{cm^{-3}}$, with the maximum droplet sizes ranging from $5 \mu$m to $1$mm. Droplets of finite size have velocities that are, in general, different from the velocity of the carrier fluid in which they are suspended. Large droplets (diameters $>100\mu$m) are dominated by gravity and undergo rapid sedimentation within range of $\sim1$m \cite{Lohse2021,Bourouiba2014}. Droplets less than $100\mu$m in size are carried away by the flow over longer distances before they settle down \cite{Bourouiba2021}. As the droplet size reduces, the response time of the droplet ($\tau_p$) gets smaller in comparison with the typical flow time ($\tau_f$). This effect is quantified by the Stokes number $St = \tau_p / \tau_f$, which indicates how important the inertial effects are for a given droplet size. For droplets with small Stokes numbers (typically less than $50\mu$m in diameter), it may be possible to coarse-grain them in the form of an Eulerian field for the liquid content (i.e., $r_l = r_l (x,y,z,t)$), while still retaining the inertial effects \cite{Ravichandran2020,ravichandran2020cumulus}. This, in essence, is the Eulerian approximation for liquid droplets which we use in this work. Such an approximation is commonly known as the one-moment scheme in the atmospheric cloud literature (see, e.g., \cite{Ravichandran2020,ravichandran2020cumulus}). When the inertial effects are retained, the velocity field of the droplets is not divergence-free, but obeys a compressible advection equation. The time rate of change of the liquid water mixing ratio, $r_l$, within this framework, can be written:
\begin{equation}
\frac{dr_l}{dt}\ \equiv \ C_d\ =\ \frac{\partial r_l}{\partial t}+\nabla\cdot(\boldsymbol{v}r_l),\label{eq:drldt-1}
\end{equation}
where, $C_d$ is condensation rate. 
The droplet velocity field ($v(t)$) can be written down using the Maxey-Riley equations \cite{maxey1983equation}
which govern the dynamics of particles in the Stokes regime:
\begin{equation}
\frac{d\boldsymbol{v}}{dt}=\frac{\boldsymbol{u}-\boldsymbol{v}}{\tau_{p}}+\boldsymbol{g},\label{eq:maxey_riley}
\end{equation}
where the droplets are assumed to be much denser than air, $\boldsymbol{g}$
is gravitational acceleration, and $\boldsymbol{u}$ is flow velocity. For small droplets, Eq. (\ref{eq:maxey_riley}) may be expanded
in powers of $\tau_{p}$ to give, to first order, 
\begin{equation}
\boldsymbol{v}=\boldsymbol{u}-\tau_{p}\frac{D\boldsymbol{u}}{Dt}+\tau_{p}\boldsymbol{g},\label{eq:maxey_approx}
\end{equation}
where $D/Dt$ is the material derivative following fluid streamlines,
and $\tau_{p}\boldsymbol{g}$ is the terminal Stokes velocity for
the droplet. Eq. (\ref{eq:maxey_approx}) can be written in non-dimensional form (choosing appropriate flow velocity and length scales) as
\begin{equation}
\boldsymbol{\tilde{v}}=\boldsymbol{\tilde{u}}+St\frac{D\boldsymbol{\tilde{u}}}{D\tilde{t}}+St\ \tilde{\boldsymbol{g}} +\mathcal{O}(St^2),\label{eq:maxey_st}
\end{equation}
where $\tilde{}$ indicates a non-dimensional quantity. In the limit of $St \to 0$, the second and third terms on the right hand side of Eq. (\ref{eq:maxey_st}) have a negligible contribution, leading to $\boldsymbol{\tilde{v}}=\boldsymbol{\tilde{u}}$. In the current study, we consider the liquid content at the orifice to represent a mono-disperse collection of $N_o$ number of droplets of diameter  $10\mu$m. This diameter is sufficiently small to expect the inertial effects to be negligible. Moreover, typical human coughs consist of a large number of droplets within the size range of $10-30\mu$m \cite{Droplet_size_I,Droplet_size_III,lindsley2013cough, yang2007size}, including some cases wherein the maximum droplet size is about $10\mu$m as mentioned earlier \cite{bourouiba2021review}. Also this is the size range which is responsible for a long-range transport of pathogens contributing to airborne transmission \cite{Rosti2020,LohsePRF2021growth}. Therefore we choose droplets of size $10\mu$m, for which the effect of the `slip' velocity between the droplets and the carrier fluid can be neglected so that the droplets effectively follow the fluid streamlines. (See a discussion in Refs. \cite{Abkarian2020} and \cite{singhal2021virus}  on this aspect with respect to speech flows.) We justify this assumption \emph{a posteriori} by computing the Stokes numbers of the droplets as a function of space and time (See figure \ref{fig_st}). For the present simulations, we take total amount of liquid expelled during the cough to be $10\mu$l \cite{Duguid1946}, giving $N_o=1.91\times10^7$ (i.e., $2.8 \times 10^4 \ \mathrm{cm^{-1}}$).
For the total cough volume of $0.679$ l, the initial volume fraction of liquid water is $1.47\times10^{-5}$, which translates into a liquid mixing ratio at the orifice of $r_{l,o}\approx 1.23\times10^{-2}$ kg/kg of dry air. During the cough (which lasts for $0.528s$), the instantaneous liquid amount expelled from the orifice is taken to be proportional to the cough velocity at that time (figure \ref{fig1}b).

\subsection{Thermodynamics of phase change \label{subsec:Thermodynamics}}

The mixing ratios of vapour and liquid water are coupled to the local temperature through the Clausius-Clapeyron law which specifies the saturation mixing ratio of vapour ($r_s$) at a given local temperature:
\begin{equation}
r_s(T) = r_{s,\infty}\ exp\left(\frac{L_v}{R_v}\left(\frac{1}{T_{\infty}}-\frac{1}{T}\right)\right) ,\label{eq:rs}
\end{equation}
where $r_{s,\infty}$ is the saturation mixing ratio of the ambient, $L_v$ is the latent heat of vapourisation of water, and $R_v$ is the gas constant for water vapour. The values for the various physical constants used in this simulation are given in table \ref{tab:table1}. As $\epsilon=\Delta T_o/T_\infty = 14/293\ \equiv  \mathcal{O}(10^{-2})$, we approximate the exponent in equation (\ref{eq:rs}) as $L_v\left(1/T_{\infty}-1/T\right)/R_v = L_v(T-T_\infty)/R_v T^2_{\infty}$, \cite{ravichandran2020cumulus}.

\begin{table}[h!]
  \begin{center}
    \caption{Physical properties used in the present study.}
    \label{tab:table1}
    \begin{tabular}{l|S|r}
      \toprule 
     \midrule 
      Latent heat of vapourisation of water\ \ \  & $L_v$ & $2.4\times10^6$\ J/kg\\
      Gas constant for water vapour & $R_v$ & $462\ $J/K kg\\
       Density of liquid water& $\rho_w$  & $1000$ kg \ m$^{-3}$ \\
       Density of dry air & $\rho_{d}$ & $1.2 $ kg \ m$^{-3}$\\
       Kinematic viscosity of air & $\nu$ & $1.5\times10^{-5}\  $m$^2$s$^{-1}$ \\
       Acceleration due to gravity& $g$  & $9.81 $ m s$^{-2}$\\
       saturation vapour mixing ratio at orifice& $r_{s,o}$ & \ \ \ $3.47\times10^{-2}$ kg/kg  of dry air\\
        saturation vapour mixing ratio in ambient \ \ \ & $r_{s,\infty}$ & $1.47\times10^{-2}$ kg/kg of dry air\\   
       \midrule %
      \bottomrule 
    \end{tabular}
  \end{center}
\end{table}

The saturation mixing ratio $r_s$ is the amount of vapour $r_v$ that can exist in equilibrium at a given temperature. In saturated parcels with $r_v > r_s$, the vapour condenses into liquid water. Conversely, the liquid water in parcels with $r_v < r_s$ evaporates. In the following, we first provide a treatment for the thermodynamics of phase change for a collection of droplets and then suggest two models that we have used for  thermodynamics of the Eulerian liquid field. For liquid droplets (which act as nuclei for condensation), the condensation and evaporation processes occur on a timescale that depends on their numbers and sizes. Note that we have implicitly assumed that the ambient has no other nuclei for condensation. The condensation rate $C_d$ can be written as
\begin{equation}
C_d \equiv \frac{dr_l}{dt}=n \cdot 4\pi a^2  \frac{da}{dt} \cdot \frac{\rho_w}{\rho_d},\label{eq:drldt}
\end{equation}
where $a$, $\rho_w$ are droplet radius and the density of liquid water respectively, and $n$ is the droplet density, i.e., number of droplets per unit cough volume. For isolated droplets growing by vapour diffusion, the droplet growth rate is given by (see, e.g., Bohren \& Albrecht \cite{bohren})
\begin{equation}
a\frac{da}{dt}= \frac{1}{C\rho_w}\left(\frac{r_v}{r_s}-1\right),\label{eq:adadt}
\end{equation}
where $C \approx 10^7$ m s kg$^{-1}$ is a weak function of temperature \cite{ravichandran2020cumulus}. The factor $r_v/r_s -1$ is the ``supersaturation''; for $r_v < r_s$, this factor is negative and the droplets shrink by evaporation. Using equation (\ref{eq:adadt}) in equation (\ref{eq:drldt}), and introducing non-dimensionalisation, we get
\begin{equation}
\frac{d \tilde{r}_l}{d \tilde{t}} = \frac{\mathcal{H}}{\tilde{\tau}_c} (\frac{\tilde{r}_v}{\tilde{r}_s}-1),\label{eq:drldt_norm}
\end{equation}
where the phase change time scale
\begin{equation}
\tilde{\tau}_c = \frac{u_c}{d_c} \frac{C \rho_d r_{s,\infty}} {4\pi a n} \label{eq:tau_c}
\end{equation}
accounts for the surface area of droplets available for evaporation, and the modified Heaviside function is given by
\begin{equation}
\mathcal{H} =
  \begin{cases} 
  1 & \text{if $r_v > r_s$ or $r_l  > 0$} \\
  0 & \text{otherwise.} \\
  \end{cases}\label{eq:H}
\end{equation}

Here, $\tilde\tau_c$ is an important parameter which tunes the rate of condensation/evaporation at any given saturation condition. Note that the quantities $a$ and $n$ appearing in the equation for $\tilde\tau_c$ (Eq. \ref{eq:tau_c}) cannot, by definition, be independently determined in a one-moment scheme and therefore further modelling needs to be done. Here we use the following two models for $\tilde\tau_c$, which depend upon on how the Eulerian liquid field is interpreted in terms of droplets.

In the first model, we determine $\tilde\tau_c$ at the beginning of the cough (at $t=0$s) and treat it as constant throughout the simulation (denoted as  `$const.\ \tau$'). This approach is typically used in cumulus cloud studies where $\tilde\tau_c$ is small and is of $\mathcal{O}(0.1)$ \cite{Ravichandran2020,hernandez2013minimal}. For the present cough flow problem, based on the initial diameter of droplets $D_o=2a_o=10\mu$m and initial droplet density $n_o=N_o/V_o$, $\tilde\tau_c$ comes out to be 14.34. This is much larger than that in clouds which is primarily due to the much smaller width of the cough flow in comparison to that of a cumulus cloud. As a result, using a constant $\tilde\tau_c$ for a cough flow is strictly not justified. Here we use it to provide a baseline case for comparison with the second model for $\tilde\tau_c$ explained below.

In the second model, we allow $\tilde\tau_c$ to be variable. Ravichandran et al. \cite{Ravichandran2020} used a variable $\tilde\tau_c$ model for computing mammatus cloud evolution, in which the number density was assumed constant in time and the droplet radius (considered mono-disperse) decreased as a function of time. Here we extend this treatment by noting that as the cough volume increases due to continued entrainment, the droplet number density goes on decreasing with time. Secondly, since the liquid content $r_l$ is a function of space and time, the product $n \times a^3$ is essentially a function of space and time (Eq. \ref{eq:m_li}). Due to the limitation of the Eulerian approximation for liquid droplets, we cannot determine both $n$ and $a$ as functions of space and time. In what follows, we will consider the number density to be a function of $t$ and the droplet radius to be a function of $x,y,z$ and $t$. This model, which represents an ``extended'' first-order moment scheme, will be denoted as  `$var.\ \tau$' and is based on the following two considerations.

\begin{enumerate}
    \item Although the droplets released from the orifice are considered mono-disperse, they are subjected to different saturation fields within the cough volume (as will be shown below). As a result one can expect the resulting droplet distribution to be poly-disperse. For example, Rosti et al. \cite{Rosti2020} have carried out a simulation with a mono-disperse droplet distribution at the orifice and observed a considerable broadening of the evaporation time caused by turbulence, implying presence of different droplet sizes.
    
    \item Since we do not consider inertial effects for liquid droplets (or liquid field), the preferential clustering of droplets will not be present \cite{Clustering_I,Clustering_II}. There can still be some concentration and dilution of liquid field as the instantaneous fluid streamlines come closer or diverge. But this effect is expected to be much weaker than the inertial clustering. It is therefore reasonable to consider an "effectively uniform" distribution of droplets so that the number density is only function of time. One may also consider this as an ``equivalent'' uniform distribution of droplets that enables (approximately) characterizing the liquid field within the cough volume as a poly-disperse distribution of droplets (considered in point 1 above).

\end{enumerate}

With these considerations, the local value of $\tilde\tau_c (\bold{x},t)$ may be written as
\begin{equation}
\tilde\tau_c (\bold{x},t) = \frac{u_c}{d_c} \frac{C\ \rho_d\ r_{s,\infty}} {4\pi\ a(\bold{x},t)\ n(t)},\label{eq:tau_ci}
\end{equation}
where $a(\bold{x},t)$ is the local droplet radius given implicitly from the definition of $r_l$ as
\begin{equation}
    \rho_d r_l(\bold{x},t) = \frac{4\pi}{3} n(t)\  a^3(\bold{x},t)\ {\rho_w}, \label{eq:m_li}
\end{equation}
and $n(t)$ is given by
\begin{equation}
    n(t) = N(t) / V(t). \label{eq:n_t}
\end{equation}
The total number of droplets (including liquid droplets and droplet nuclei) in the cough volume, $N(t) = N_o$ for $t > t_o=0.528$s.
For $t < t_o$, $N(t)$ is taken to be proportional to the fluid volume expelled ($=\int^t_0$ `cough flow-rate' $dt$) at the origin up to time $t$. $V(t)$ is the volume of cough at time $t$ and is determined based on a threshold for the cough-flow presented in section \ref{sec:results}.

\subsection{Governing equations} \label{sec:nondim_eqns}
As the flow velocities and the temperature differences in the flow are small, we make the Boussinesq approximation, with density differences appearing only in the buoyancy term. This approximation has been used in the recent DNS and LES studies on cough and speech flows \cite{Lohse2021,Rosti2020,Abkarian2020,singhal2021virus}. The equations governing the dynamics are thus the incompressible Navier-Stokes equations for the velocity, coupled with scalar equations for the temperature and mixing ratios of vapour and liquid water with appropriate source/sink terms for the scalars.
\begin{align}
\nabla\cdot\boldsymbol{u} &= 0; \label{eq:continuity-1}\\
\frac{D\boldsymbol{u}}{Dt} &= -\frac{\nabla p}{\rho_\infty}+\nu\nabla^{2}\boldsymbol{u}+\boldsymbol{B};\label{eq:momentum-1}\\ 
C_p\frac{D\theta}{Dt} &=  \kappa\nabla^{2}\theta+L_{v}C_{d};\label{eq:temperature-1}\\
\frac{Dr_v}{Dt} &=   \kappa_v\nabla^{2}r_v - C_{d};\label{eq:vapour-1}\\
\frac{Dr_l}{Dt} &=  \kappa_l\nabla^{2}r_l + C_{d}.\label{eq:liquid-1}
\end{align}
Here, 
$\nu$ is the fluid viscosity of air, $\kappa$, $\kappa_{v}$ and $\kappa_{l}$ are the diffusivities of temperature, vapour and liquid, respectively, $L_v$ is the latent heat of vaporisation of water, and $\theta = T - T_{\infty}$ is the temperature difference. The buoyancy term, obtained within the Boussinsq approximation, is given by,
\begin{equation}
    \boldsymbol{B}=g\frac{\rho_{\infty}-\rho}{\rho_{\infty}}\hat{e}_y=g\left[\frac{T-T_{\infty}}{T_{\infty}}+(\xi-1)(r_{v}-r_{v,\infty})-r_{l}\right]\hat{e}_y,\label{eq:buoyancy}
\end{equation}
where $\xi$ ($=1.61$) is the ratio of gas constants of air and water vapour. This is obtained by writing $\rho = \rho_d (1 + r_v + r_l)$ and $\rho_{\infty} = \rho_{d,\infty} (1+r_{v,\infty})$, and by performing a linearisation for small magnitudes of $r_l$ and $r_v$. A detailed derivation of Eq. (\ref{eq:buoyancy}) can be found in Ravichandran and Narasimha \cite{Ravichandran2020}.

Equations (\ref{eq:continuity-1}--\ref{eq:liquid-1}) are nondimensionalised using the length scale $d_c$, the velocity scale $u_c$, the temperature scale $\Delta T_o$, and the scale for water mixing ratios $r_{s,\infty}$, giving the nondimensional governing equations

\begin{align}
\tilde{\nabla}\cdot\tilde{\boldsymbol{u}} & =\  0,\label{eq:continuity}\\
\frac{D\tilde{\boldsymbol{u}}}{D\tilde{t}} & =  -\tilde{\nabla} \tilde{p} +\frac{1}{Re} \tilde{\nabla}^{2}\tilde{\boldsymbol{u}}+
\frac{1}{Fr^{2}}\left[ \tilde{\theta}+ \frac{r_{s,\infty}}{\epsilon} \left((\xi-1) (\tilde{r}_{v}-\tilde{r}_{v,\infty})-\tilde{r}_{l}\right)\right]\hat{e}_{y},\label{eq:momentum}\\
\frac{D\tilde{\theta}}{D\tilde{t}} & =\frac{1}{RePr}\tilde{\nabla}^{2}\tilde{\theta}+L_{1}\tilde{C}_{d},\label{eq:temperature}\\
\frac{D\tilde{r}_{v}}{D\tilde{t}} & =\frac{1}{ReSc_{v}}\tilde{\nabla}^{2}\tilde{r}_v-\tilde{C}_{d},\label{eq:vapour}\\
\frac{D\tilde{r}_{l}}{D\tilde{t}} & =\frac{1}{ReSc_{l}}\tilde{\nabla}^{2}\tilde{r}_l+\tilde{C}_{d}\label{eq:liquid},
\end{align}

where the Reynolds number $Re=d_c u_c/\nu = 4849$, the inverse-square Froude number $Fr^{-2}=g \epsilon d_c/u^2_c = 1.01\times 10^{-3}$, the Prandtl number $Pr=0.71$, the Schmidt numbers $Sc_{v,l}=1$, and the nondimensional latent heat of vaporisation
\begin{equation}
    {L}_1= \frac{L_v r_{s, \infty}}{C_p T_\infty \epsilon} =  2.73.\label{eq:l1}
\end{equation}
In above equations (\ref{eq:continuity}--\ref{eq:liquid}), all the variables with $\tilde{}\ $s represent normalised quantities of their respective parameters, for example, $\tilde{\boldsymbol{u}}$ represents the normalised velocity vector.
The saturation vapour mixing ratio in Eq. \ref{eq:rs} is also normalised and will be used in a form given as,
\begin{equation}
    \tilde{r}_s(\tilde{\theta})= exp(L_2 \tilde{\theta}), \label{eq:norm_rs}
\end{equation}
where nondimensional constant $L_2$ is,
\begin{equation}
    {L}_2= \frac{L_v \epsilon}{R_v T_\infty }. \label{eq:l2}
\end{equation}

\subsection{Numerical method and code validation \label{subsec:Numerical-method}}

\begin{figure}[!h]
\centering%
\begin{minipage}{0.46\textwidth}
  \includegraphics[width=1.0\textwidth]{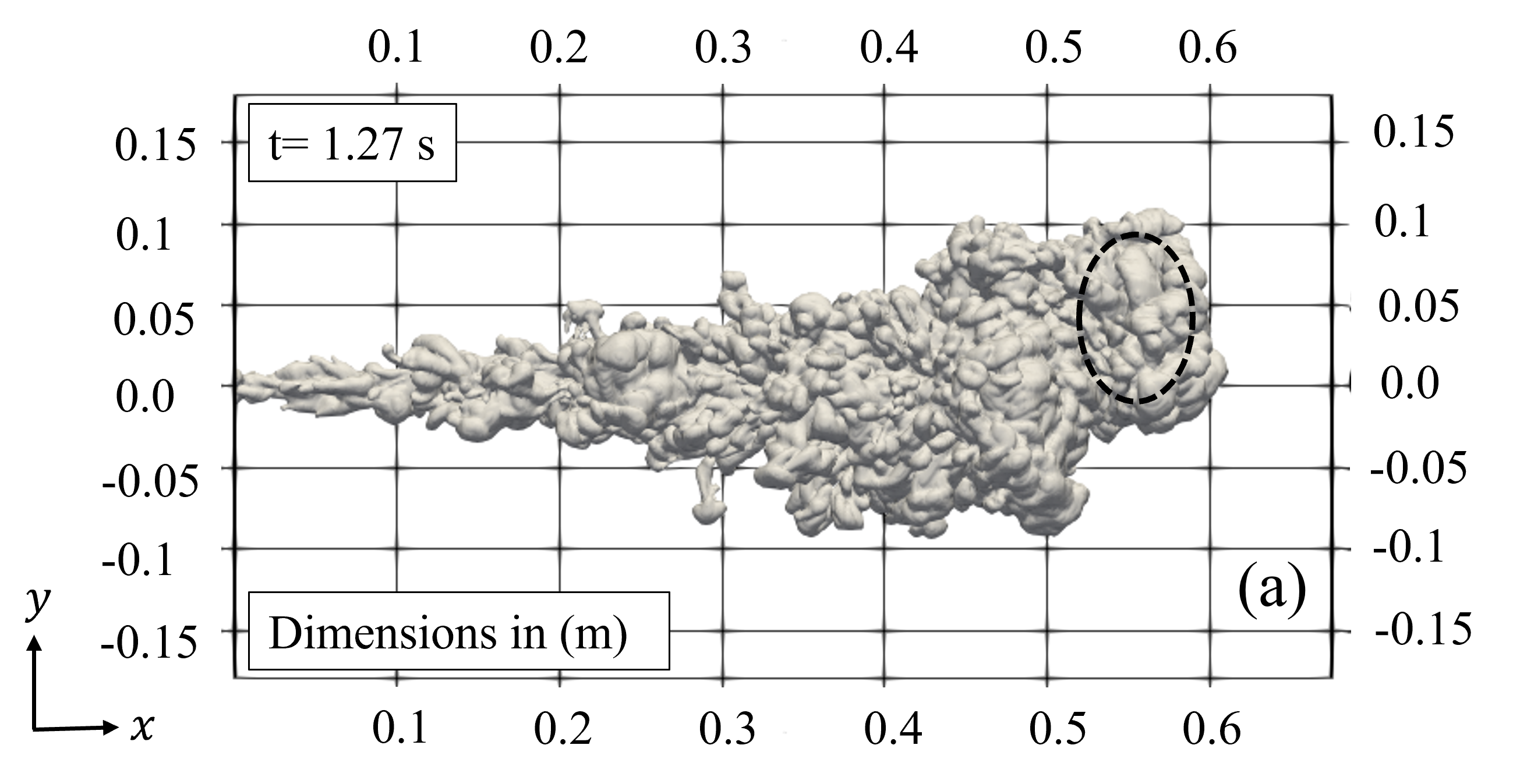}
\end{minipage}%
\hspace{0.00\textwidth}%
\begin{minipage}{0.52\textwidth}
  \includegraphics[width=1.0\textwidth]{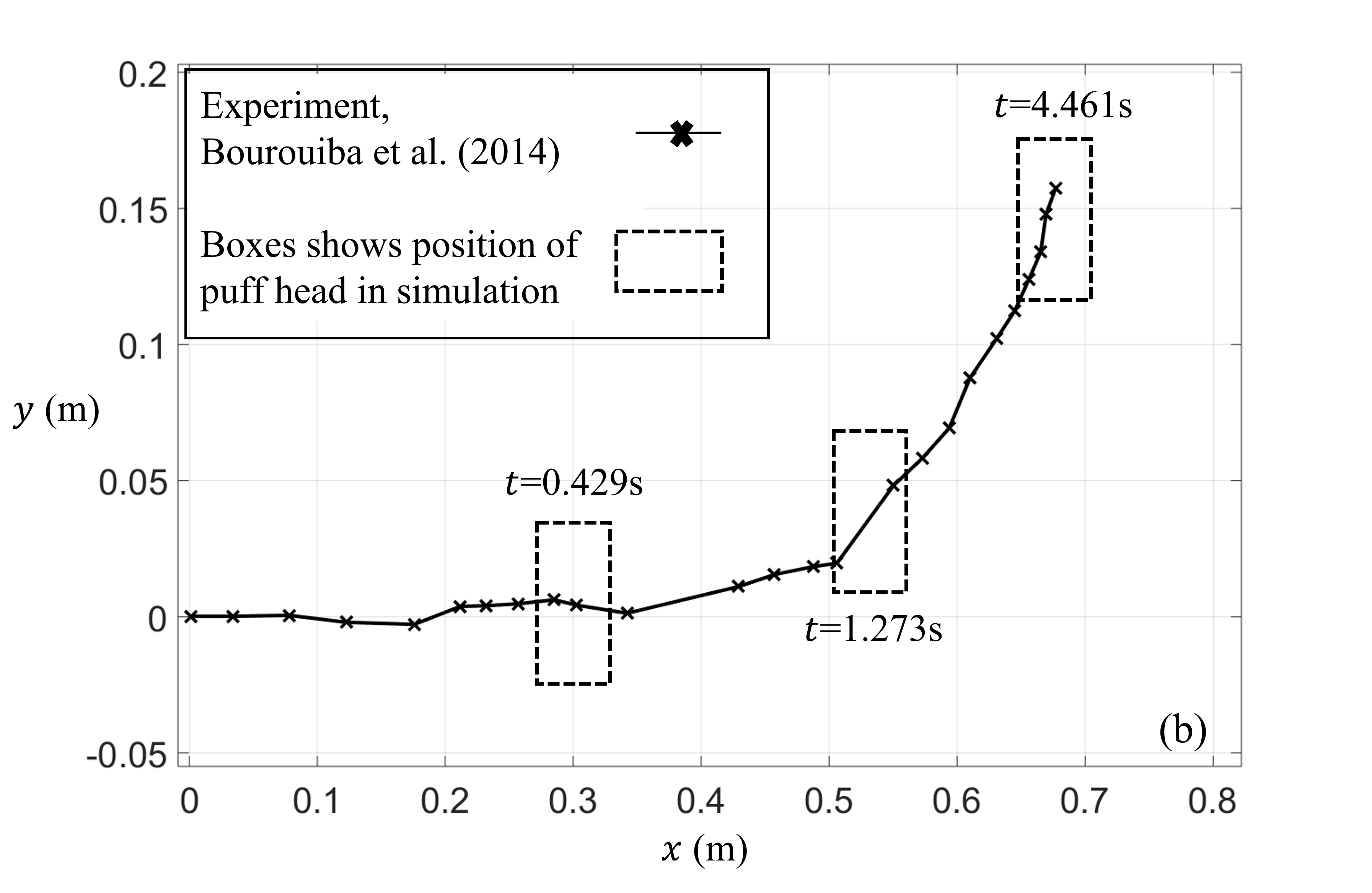}
\end{minipage}
\caption{(a) Iso-surface of the density difference at 1\% of the  orifice value at a time $t=1.27$s after flow initiation. The dashed circle representing the ``head'' of the flow is drawn based on the front edge and shape of this iso-surface. (b) The trajectory of the ``head'' of the flow from (a) is overlaid on the trajectory of a cough flow from the experimental measurements of \cite{Bourouiba2014} (see their figure 11).}
\label{fig2}
\end{figure}

Equations (\ref{eq:continuity}-\ref{eq:liquid}) are solved in a cuboidal domain of dimensions $80d (=1.8$m in $x$-direction$) \times 40d \times 40d$ using the finite-volume solver \emph{Megha}-5, which is second-order accurate in space and time. The domain is discretised with uniform and equal grid spacing ($\Delta$) in all three directions, with a total of $1024 \times 512 \times 512$ grid points. 
The grid resolution in the present study is found to be as good as or better than that reported in the literature on cough-flow DNS \cite{Lohse2021,Rosti2020}; see section S1 in the supplementary material. 
A second-order Adams-Bashforth scheme is used for time-stepping, with a CFL number of 0.15. Convective boundary conditions are imposed on all flow variables at the $x=L_x$, $y=\pm L_y/2$ and $z=\pm L_z/2$ boundaries (see figure~\ref{fig1}a). A free-slip, no penetration boundary condition on the velocity, and zero-flux boundary conditions on the scalars are imposed at $x=0$, except for the orifice where Dirichlet boundary conditions are specified. The solver has been extensively validated and has previously been used to study the statistics of steady jets \& plumes \cite{Singhal2021}, cumulus \& mammatus clouds \cite{Ravichandran2020}, as well as previous studies of virus transmission by respiratory flows \cite{singhal2021virus,diwan2020understanding}.
Each of the present simulated cases was run on the CRAY supercomputer (CRAY XC40) using 2048 number of cores and the total run time of about $43000$ core hours.

As a further validation test, we simulate the ``dry'' cough case (i.e. a puff which is lighter than the ambient, but does not contain evaporating water droplets), with the experimental results from Bourouiba et al. \cite{Bourouiba2014} for the ``Case I'' in their experiment (see table 1 in Ref. \cite{Bourouiba2014} ).  In this experimental case, Bourouiba et al. \cite{Bourouiba2014} injected saline payload of $88$cm$^3$ in a water tank with a density difference  of $3.15\times10^{-3}$g/cm$^3$, treated as a buoyancy scalar which under Boussinesq approximation acts as a source term for the vertical velocity; the fluid velocity at the orifice was kept approximately constant during the release of payload. We have replicated these conditions in our simulation for comparison. The simulated cough flow at time $t=1.27$s, visualized by an iso-surface of density difference, is shown in figure \ref{fig2}a. The trajectory of the ``head'' of this dry cough obtained numerically is compared with the experimental result in figure \ref{fig2}b, and the two results can be seen to agree well. This provides an additional support to the buoyancy module of the code. 

\section{Results} \label{sec:results}
 Here, we present simulation results for the mild cough flow described in section \ref{subsec:geometry} using the Eulerian models detailed in section \ref{subsec:Thermodynamics}. We first discuss how the closure assumption of uniform (in space) number density of droplets, Eq. (\ref{eq:tau_ci}), affects the dynamics in ``$var. \tau$'' model and compare the results between the two models. In order to specify the droplet number density in Eq. (\ref{eq:tau_ci}) for the $var. \tau$ model, the volume of the cough $V(t)$ at each time instant needs to be be defined in a self-consistent manner; i.e., the turbulent puff with a complex boundary has to be delineated from the ambient. We define the cough volume as the volume of the flow with vapour mixing ratio larger than a chosen threshold. Once $V(t)$ is determined at a given time instant, $n(t)$ can be calculated from Eq. (\ref{eq:n_t}) and $a(x,t)$ (droplet radius) from Eq. (\ref{eq:m_li}), from which $\tilde{\tau}_c$ is obtained (Eq. \ref{eq:tau_ci}). This enables calculating the condensation/evaporation rate $\tilde{C}_d (t)$ which is used to solve the equations for the next time instant (Eqs. \ref{eq:temperature}, \ref{eq:vapour}, \ref{eq:liquid}). Thus the threshold used for determining $V(t)$ needs to be ``pre-set'' into the code for marching the solution in time. 
 
 \begin{figure}[!h]
\centering%
\begin{minipage}{0.51\textwidth}
  \includegraphics[width=1.0\textwidth]{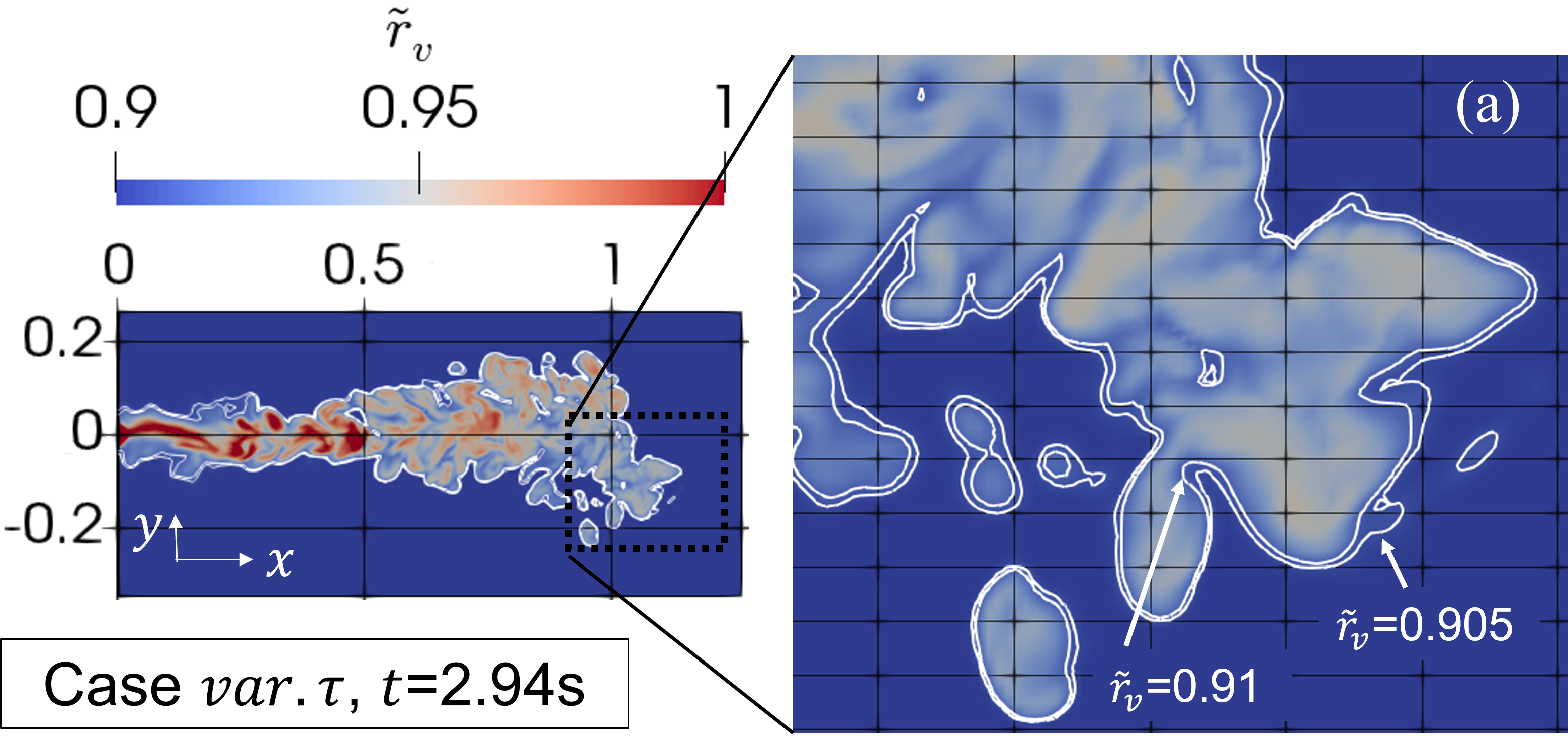}
\end{minipage}%
\hspace{0.00\textwidth}%
\begin{minipage}{0.49\textwidth}
  \includegraphics[width=1.0\textwidth]{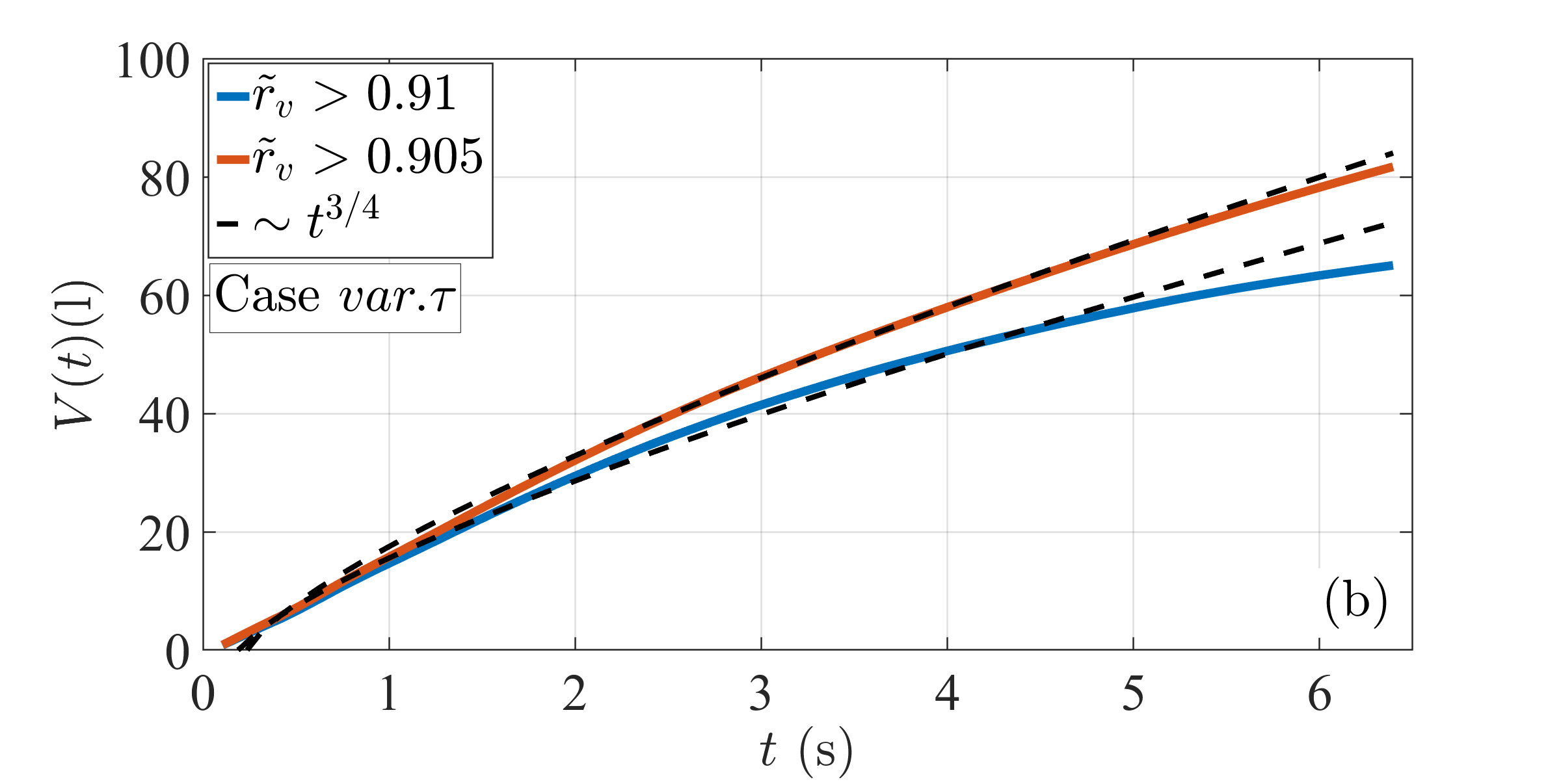}
\end{minipage}
\caption{ (a) Filled contours for the vapour mixing ratio $\tilde r_v$ overlaid with line contours (in white) corresponding to $\tilde r_v = 0.905$ and  $\tilde r_v = 0.91$ in the vertical ($z=0$) plane. The inset shows a zoomed-in view at the location indicated, showing that the line contours do not change significantly as the threshold value is changed. (b) Cough volume $V(t)$ based on the two thresholds shown in (a); $V(t)$ variation approximately obeys the $3/4$ scaling for a turbulent puff.}
\label{fig3}
\end{figure}
 
 Figure \ref{fig3}a shows an instantaneous distribution of $\tilde{r}_{v}$ with two line contours corresponding to the thresholds of $\tilde{r}_{v}=0.905$ and $\tilde{r}_{v}=0.91$. As can be seen, both the thresholds are effective in delineating the cough flow from the ambient. The cough volumes calculated based on these two thresholds are plotted as a function of time in figure \ref{fig3}b. There is a considerable increase in $V(t)$ with time compared to the orifice cough volume of $0.679$ l; this is due to the turbulent entrainment of the ambient fluid into the cough flow \cite{diwan2020understanding}. At a given time instant, $V(t)$ for the threshold of $\tilde{r}_{v}=0.905$ is larger than that for $\tilde{r}_{v}=0.91$ as expected, and the difference between them increases to about $15\%$ after 6s (figure \ref{fig3}b). Thus the precise choice of the threshold will affect the droplet number density (Eq. \ref{eq:n_t}) and therefore the value of $\tilde{\tau}_c$. However, for a given $r_l$, an increase (decrease) in the number density (due to a different choice of the threshold) causes a decrease (increase) in the droplet radii (Eq. \ref{eq:m_li}), which has a compensating effect for the determination of $\tilde{\tau}_c$ (Eq. \ref{eq:tau_ci}). As a result, the thermodynamic quantities like the total liquid content and the net evaporation rate do not show much difference with a change in the threshold value; see section S2 in the supplementary material for more details. This situation is entirely acceptable considering the scope of the $var. \tau$ model, which is supported by a good comparison of our results (presented below) with the available literature.  In what follows, we choose a threshold of $\tilde{r}_{v}=0.91$. Note that the cough volume follows the expected $V(t)\sim t^{3/4}$ variation for both the values of threshold as seen in figure \ref{fig3}b consistent with the previous results (see, e.g., \cite{Rosti2020}).

\subsection{Comparison of results between the two models for $\tilde{\tau}_c$}

\begin{figure}[!h]
\centering%
\begin{minipage}{0.49\textwidth}
  \includegraphics[width=1.0\textwidth]{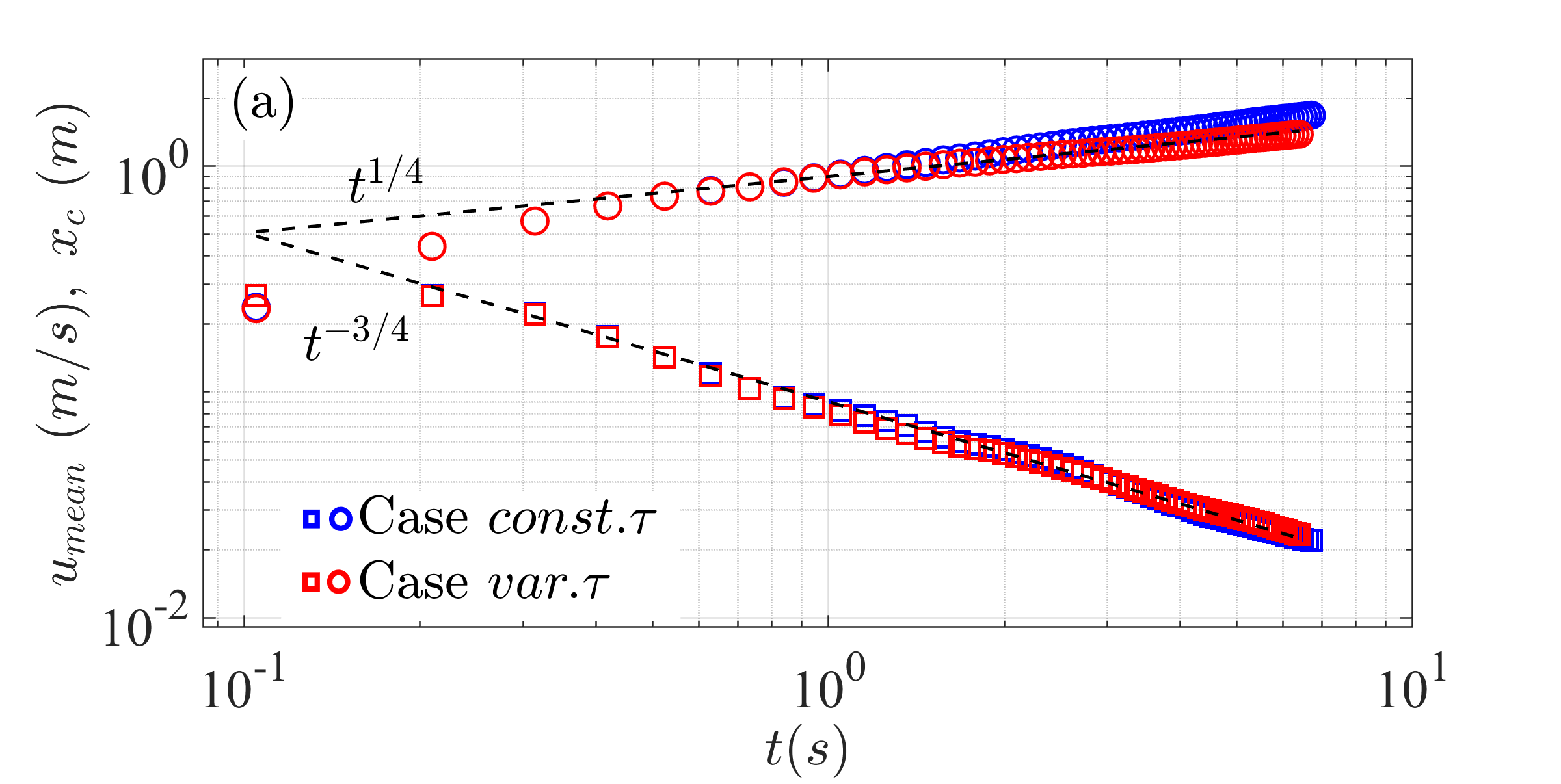}
\end{minipage}%
\hspace{0.01\textwidth}%
\begin{minipage}{0.49\textwidth}
  \includegraphics[width=1.0\textwidth]{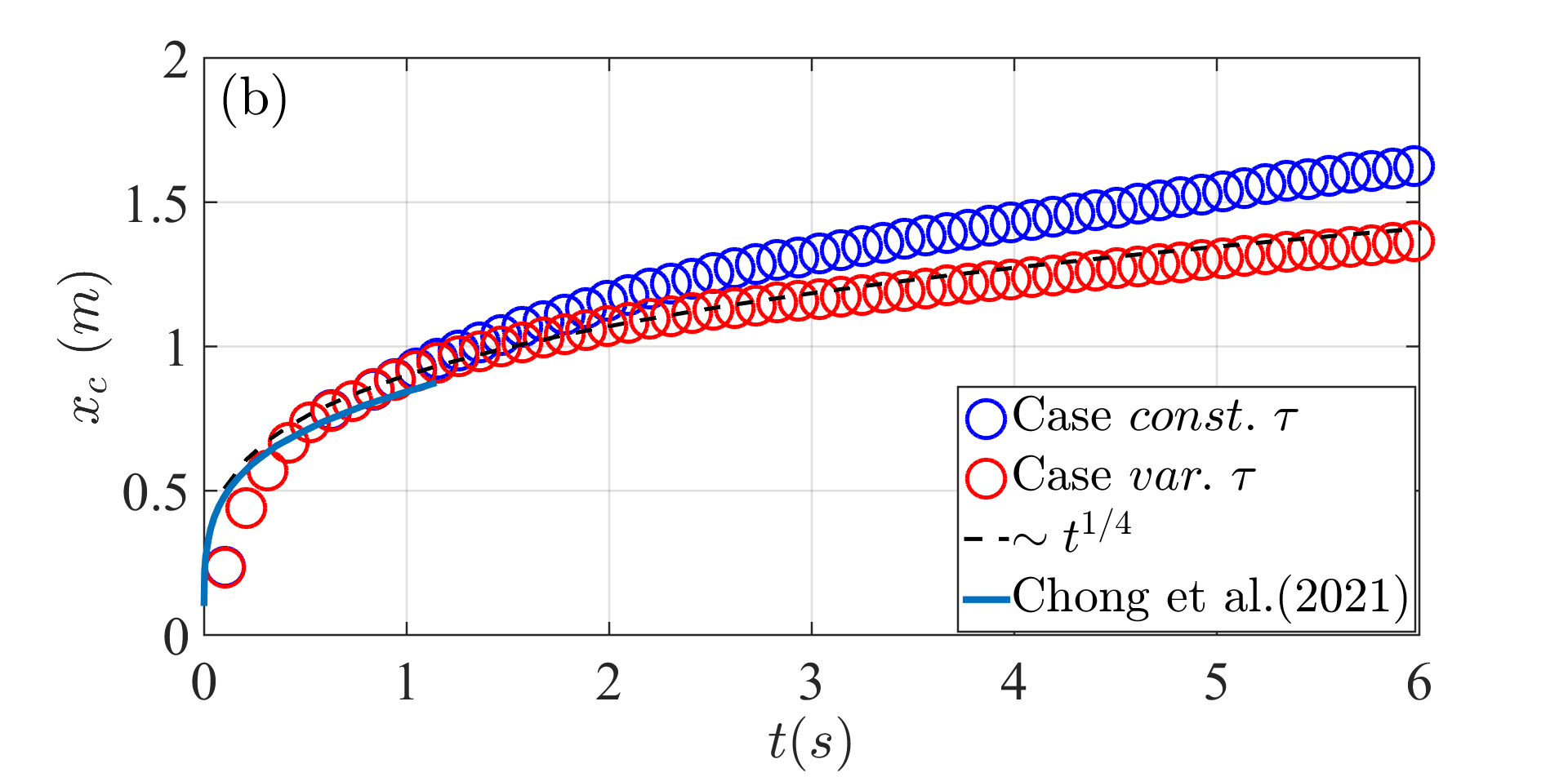}
\end{minipage}
\caption{(a) Mean velocity $u_{mean}$ (squares) and streamwise extent (or horizontal ``reach'') $x_c$ (circles) of the cough flow as a function of time. The dashed lines show the expected behaviour for a turbulent puff. The colours indicate the two models for $\tau_s$ (see section \ref{subsec:Thermodynamics}). Note that the axes here are logarithmic. (b) Time variation of $x_c$ plotted on linear axes.}
\label{fig_velo_reach}
\end{figure}

We start by comparing broad features of the cough flow for the $const.\tau$ and $var.\tau$ models. The mean velocity of the cough flow ($u_{mean}$) and its streamwise extent ($x_c$) are plotted in figure \ref{fig_velo_reach}a for the two models. $u_{mean}$ is calculated by averaging the streamwise component of velocity over the cough volume $V(t)$ at a given time instant and $x_c$ is determined as the farther point from the orifice where $\tilde{r_v}>0.91$. The exhaled total momentum of the cough flow in the streamwise direction can be expected to be constant. Therefore, the cough-flow velocity should decrease with time following $t^{-3/4}$, as the puff volume increases with the power law $t^{3/4}$ (figure \ref{fig3}b). Figure \ref{fig_velo_reach}a shows that the mean cough-flow velocity follows the $t^{-3/4}$ relation well, for both the models. The streamwise extent or ``reach'' of the cough flow is another important parameter, as it determines how far droplets from an infected person can potentially transmit the virus.
As seen from figure \ref{fig_velo_reach}a, $x_c$ follows the $t^{1/4}$ variation (i.e., cube-root of $V(t)$) well for the $var.\tau$ model, whereas the $const.\tau$ model shows a continuous departure from this law as time increases. This is seen more clearly in the linear graph for $x_c$ shown in figure \ref{fig_velo_reach}b, wherein the $const.\tau$ model is seen to depart from the $t^{1/4}$ variation for $t>2$s. Interestingly, this departure for the $const.\tau$ model is related to the size and behaviour of a chunk of the cough-flow which separates from the main flow near its head. This feature will be discussed in relation to figure \ref{fig_axial_liquid}(a) below; see also figure S4(a) in the supplementary material. Figure \ref{fig_velo_reach}b also shows a comparison of the present results with those reported by Chong et al. \cite{Lohse2021}. As can be seen, there is a good match between the $x_c$ variation from Ref. \cite{Lohse2021} and that for the $var.\tau$ model from the present study.

\begin{figure}[!h]
\centering%
\begin{minipage}{0.60\textwidth}
  \includegraphics[width=1.0\textwidth]{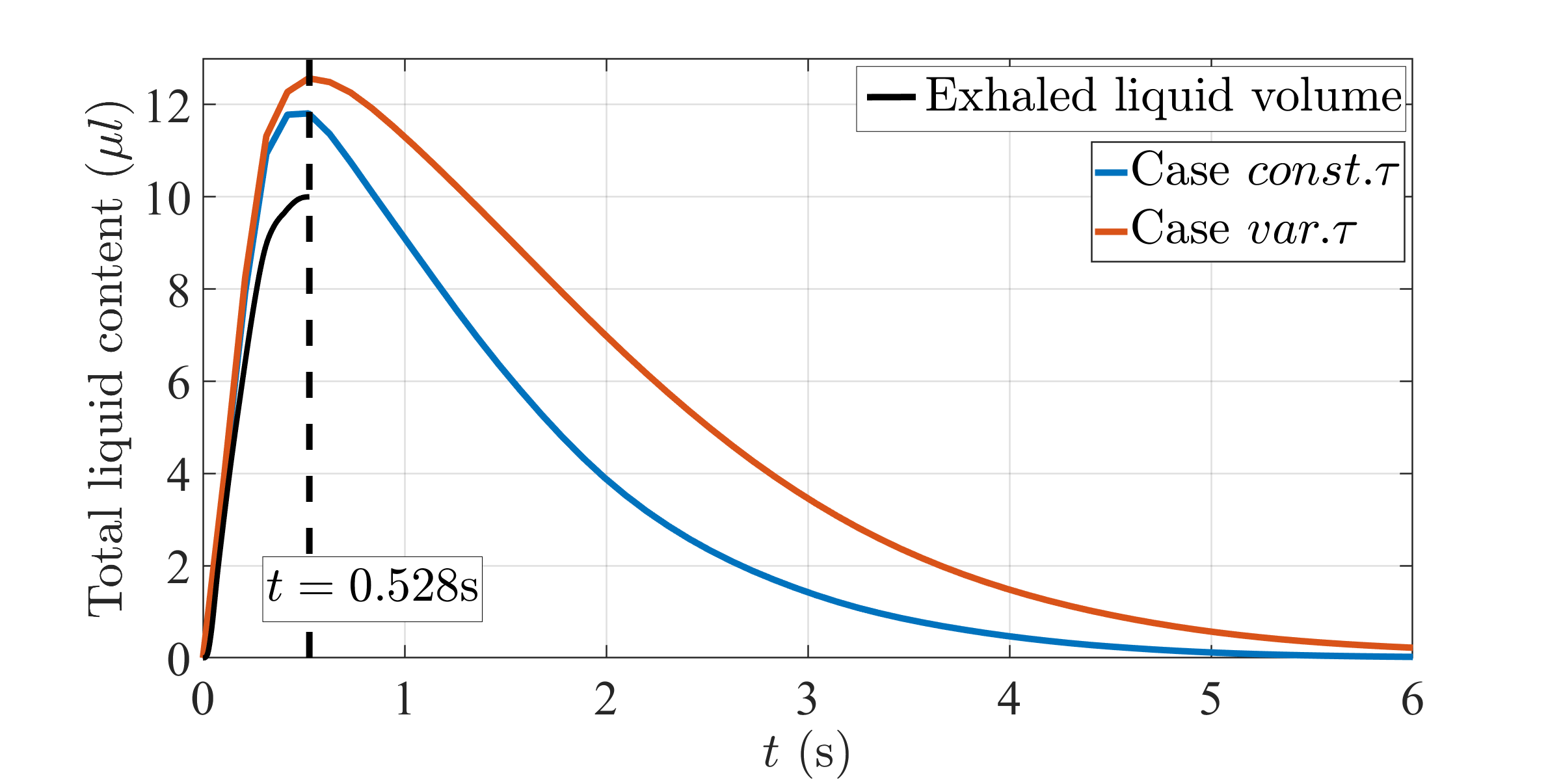}
\end{minipage}%
\caption{Time variation of the total liquid content in the cough flow for the $const. \tau$ and $var. \tau$ models.}
\label{fig_liq_cont}
\end{figure}

Figure \ref{fig_liq_cont} shows the total liquid-water content within the cough volume as a function of time. The black solid line indicates the liquid volume exhaled at the orifice; a total of $10\mu$l is expelled over the cough duration of $0.528$s. The total liquid contents for both the models show an initial increase and exhibit consistently larger values than the exhaled liquid content (figure \ref{fig_liq_cont}). At the end of the cough duration, there is an increment in liquid volume of 18\% and 25\% respectively for the $const. \tau$ and $var. \tau$ models compared to the total exhaled liquid. This implies that there is an initial condensation of the water vapour expelled from the orifice, possibly due to supersaturation, causing an increase in the liquid volume. In this connection, Chong et al.\cite{Lohse2021} found that higher saturation and lower temperature conditions of the ambient (90\% and $20^\circ $C) were favorable for longer survival of droplets in their simulations, again due to condensation of the supersaturated water vapour in the cough. We have replicated the same ambient conditions here and therefore the trend in figure \ref{fig_liq_cont} indicates that we have been able to accurately capture the physical effects of supersaturation using the present scheme. After  $t=0.528$s, the liquid volume starts decreasing as the droplets evaporate in the sub-saturated environment created due to the dilution of the cough flow by the entrained ambient air (figure \ref{fig_liq_cont}). The evolution of the saturation field within the cough flow will be presented in some detail in section \ref{subsec:main_results}.

Comparing the $const. \tau$ and $var. \tau$ models in relation to figure \ref{fig_liq_cont} shows that the $const. \tau$ model considerably under-predicts the liquid content as compared to that for the  $var. \tau$ model at a given time instant (especially for $t>0.528$). Alternatively, liquid water takes a longer time to evaporate down to a specified level for the $var. \tau$ model compared to that for the  $const. \tau$ model. This is due to an increase in the value of $\tilde{\tau}_c$ (implying longer evaporation time scales) for the former model as a result of a decrease in the number density and droplet size with time (Eq. \ref{eq:tau_ci}); for the latter model $\tilde{\tau}_c$ is kept fixed giving a constant evaporation time scale. We therefore expect the $var. \tau$ model to be more realistic in simulating the thermodynamics of phase change in mild cough flows and present more results for this case in the next section. A few additional results for the $const. \tau$ model are presented in figures S3 and S4 in the supplementary material. 

Figure \ref{fig_liq_cont} gives us useful information about the time it takes for most of the expelled liquid to get evaporated. We find that it takes about 10 times the cough duration for the liquid content to drop to less than 5\% of its initial value for the $var. \tau$ model (and 7.5 times the cough duration for the $const. \tau$ model;  figure \ref{fig_liq_cont}). During this time, the streamwise distance travelled by the cough flow is about 1.3-1.5m (figure \ref{fig_velo_reach}b). Liquid droplets in such flows are thus long-lived. (We note, however, that we have assumed an initially monodisperse droplet size distribution.) It is expected that for a stronger cough the lifetimes and the distances travelled would be even larger.

\subsection{Flow evolution and thermodynamics for the $var. \tau$ model}\label{subsec:main_results}

\begin{figure}[!h]
\centering%
\begin{minipage}{0.49\textwidth}
  \includegraphics[width=1.0\textwidth]{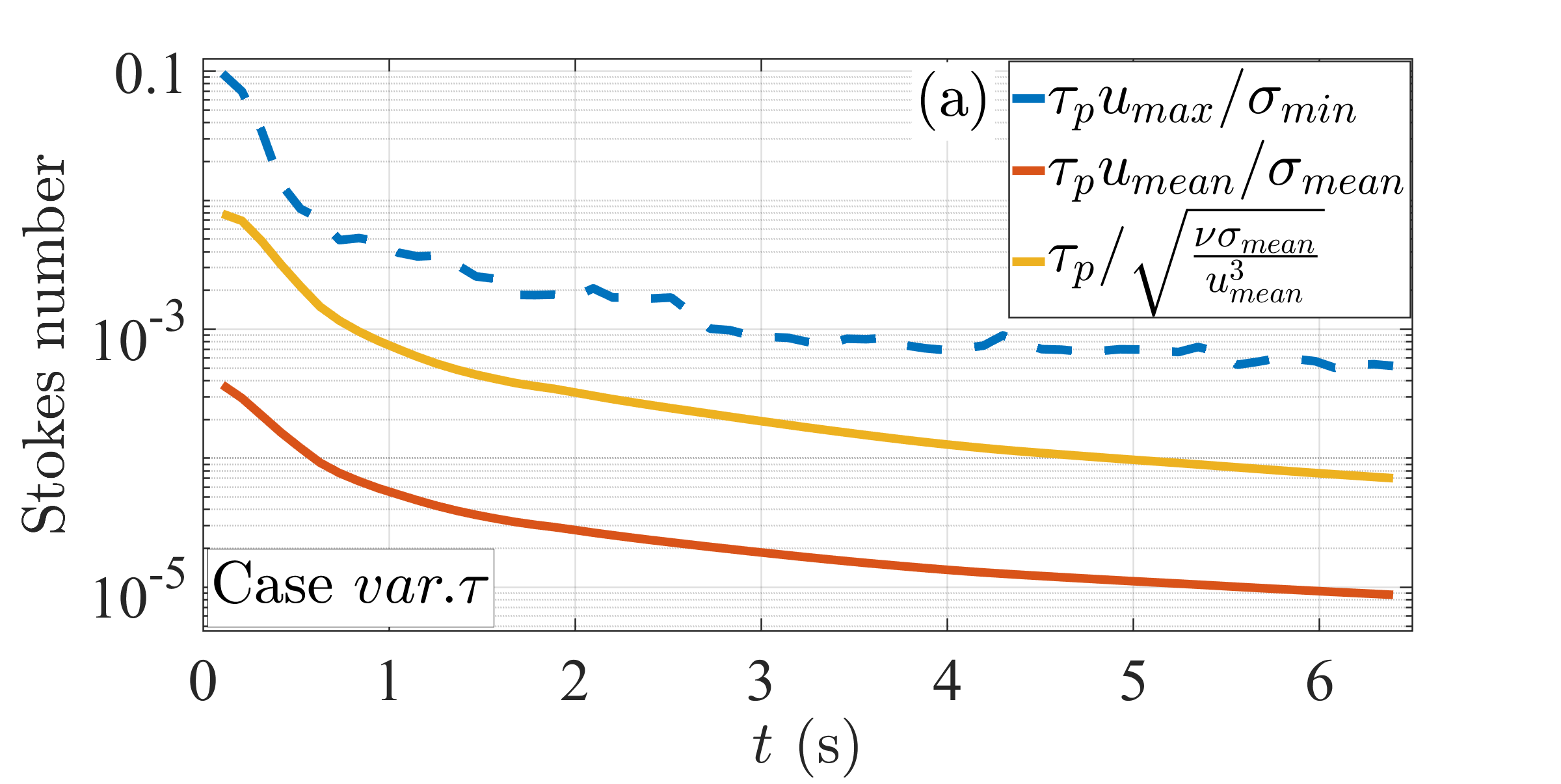}
\end{minipage}%
\hspace{0.01\textwidth}%
\begin{minipage}{0.49\textwidth}
  \includegraphics[width=1.0\textwidth]{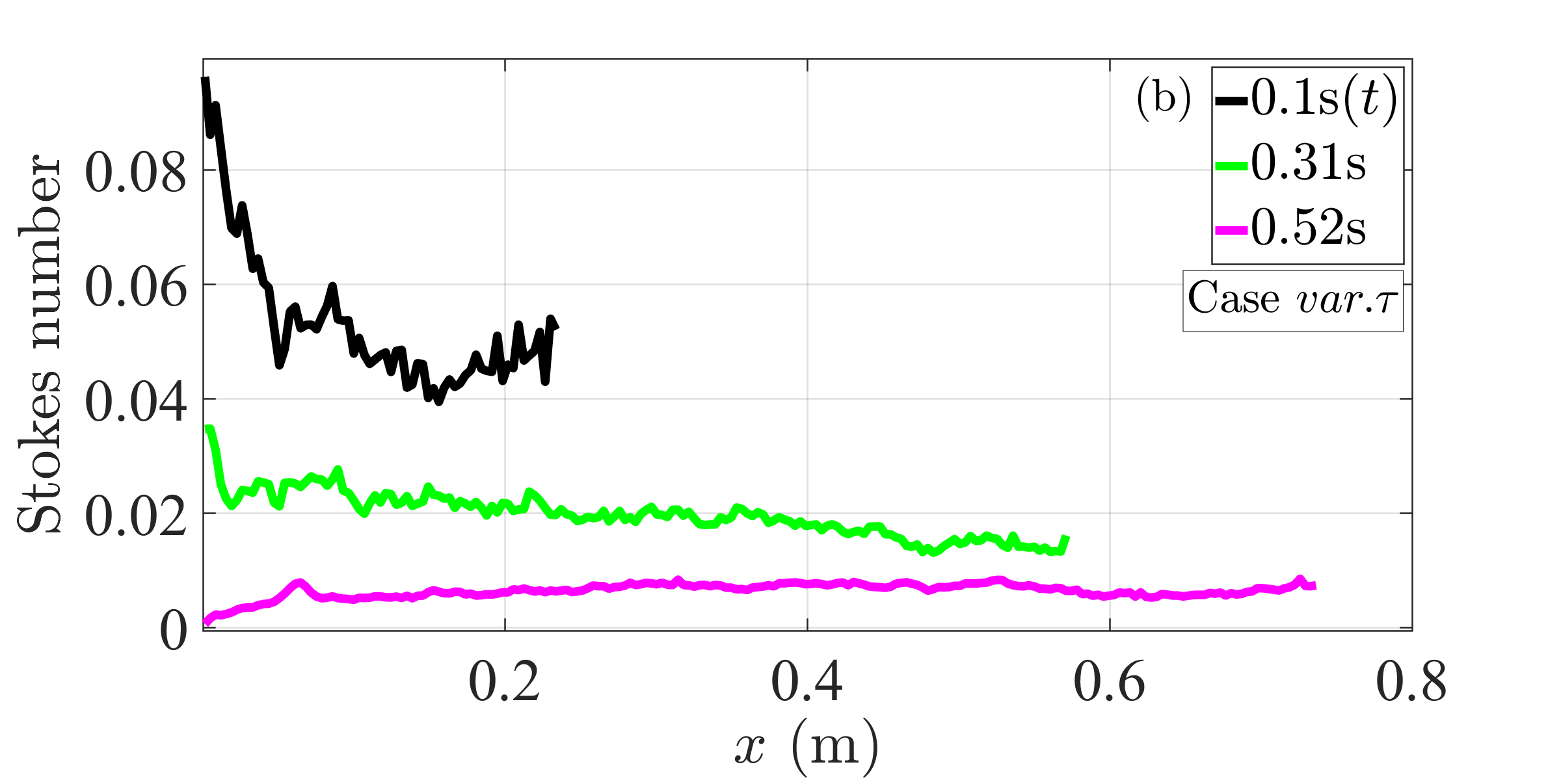}
\end{minipage}%
\caption{(a) Variation of the Stokes number for the cough flow ($St=\tau_p / \tau_f$) with time, where $\tau_p$ and $\tau_f$ are the droplet-response and flow time scales respectively, with three different measures for $\tau_f$. The initial droplet diameter is $2a_o=10\mu m$. (b) Streamwise variation of the local Stokes number for three time instants.}
\label{fig_st}
\end{figure}

Figure \ref{fig_st}a shows the variation of the Stokes number for the cough flow, $St=\tau_p/\tau_f$, as a function of time. The droplet response (or relaxation) time is calculated as $\tau_p=\rho_l D^2_o/18\mu_g$, for the initial droplet diameter ($D_o = 2a_o$) of  $10 \mu$m; here $\mu_g$ is the dynamic viscosity of air. The flow time scale $\tau_f$ is typically written as the ratio of a flow length scale to a flow velocity scale. We use three different measures for $\tau_f$ based on different choices of velocity and time scales. A natural choice is to use the \textit{mean} cough-flow velocity $u_{mean}$ (averaged over $V(t)$; see figure \ref{fig_velo_reach}a) and the \textit{mean} width of the cough-flow $\sigma_{mean}$ defined in Eq. (\ref{eq:sigma_mean}). 
\begin{equation}
\frac{\pi}{4} \sigma^2_{mean}=   mean\left[\int_{y,z} dy dz \right] \ \ \ \ for\ \  \tilde{r}_v>0.91. \label{eq:sigma_mean}
\end{equation}
For this measure of $\tau_f$, which represents a large-eddy turnover time, the Stokes number is found to be negligibly small, dropping from about $5 \times 10^{-4}$ to less than $10^{-5}$ as time progresses (figure \ref{fig_st}a). The second choice for $\tau_f$ is taken to be the Kolmogorov time scale, which is the smallest time scale in the turbulence cascade, given as $\sqrt{\nu \sigma_{mean}/u^3_{mean}}$ \cite{Rosti2020}. This is based on the estimate of energy dissipation as $\sim u^3_{mean}/\sigma_{mean}$. The Stokes number calculated from this measure of $\tau_f$
shows somewhat higher values than that based on the large-eddy time but the absolute values of $St$ are still fairly small, ranging from $10^{-2}$ to $10^{-4}$ (figure \ref{fig_st}a).

The third measure of $\tau_f$ is obtained based on the maximum velocity ($u_{max}$) and the minimum flow width ($\sigma_{min}$) of the cough flow as the velocity and length scales; $\sigma_{min}$ is defined in Eq. (\ref{eq:sigma_min}). This results in the maximum possible values for the Stokes number at a given time instant, shown in figure \ref{fig_st}a as dashed lines. Note that these  $St$ values are not necessarily realized by the cough liquid droplets but represent an upper bound which is not likely to be exceeded by any droplet. As seen from figure \ref{fig_st}a, the upper bound on $St$ has a value of $10^{-1}$ at $t=0$ but drops rapidly to $10^{-2}$ after the end of cough duration ($\sim 0.5)$s and continues to drop to reach values less than $10^{-3}$ at $t=6$s. Thus even the largest possible estimates of $St$ within the cough flow are less than $10^{-1}$. 
\begin{equation}
\frac{\pi}{4} \sigma^2_{min}=  \underset{x}{min}\left[\int_{y,z} dy dz \right] \ \ \ \ for\ \  \tilde{r}_v>0.91. \label{eq:sigma_min}
\end{equation}

This exercise shows that for the droplets of the order of $10 \mu$m in diameter and with flow parameters representing a mild cough, the Stokes numbers within the flow domain are sufficiently small ($<<1$) for the droplet inertia to be negligible. This provides a support to our earlier premise (section \ref{subsec:tracer}) that the droplets in our simulation follow the streamlines of the flow and that the effect of slip velocity can be neglected. Note that even for a somewhat larger droplet diameter of, say, $30\mu$m, the Stokes numbers based on the Kolmogorov time scale can be expected to be of the order of  $10^{-2}-10^{-3}$, which are $<<1$. Thus the present formulation can, in principle, be applied to somewhat larger droplets sizes as well.

Figure \ref{fig_st}b shows the streamwise variation of the Stokes number at three time instants of the cough duration: $0.1$s  which is close the peak cough flow rate, $0.52$s which is close to the end of the cough duration, and $0.31$s which is an intermediate time instant (figure \ref{fig1}b). For these cases, the local $\tau_f$ (i.e., at a given $x$) is calculated based on the local width of the flow and the maximum streamwise velocity at that $x$. At $t=0.1$s, $St$ values are somewhat higher ($6\times 10^{-2} - 10^{-1}$) for $x<0.05$m but beyond this distance they are smaller. For the two later time instants $St$ is typically of the order of $10^{-2}$, providing a further support to our main premise. In this connection, it is relevant to refer to the results of Rosti et al. \cite{Rosti2020} who carried out a numerical experiment for an initial mono-disperse droplets of $10\mu$m diameter, wherein simulations were carried out with and without droplet inertia. They plotted the centre of mass of the cloud of droplets (figure  9 in Ref. \cite{Rosti2020}) and found that the $x$ location of the centre of mass was closer to the orifice when inertia was accounted for as compared to the no-inertia case. This can be understood in relation to figure \ref{fig_st}b ($t=0.1$s and $0.31$s), wherein droplets with higher Stokes numbers are found located closer to the orifice. If such droplets are large enough for the inertial effects to be important (say, due to much higher $Re$ and therefore smaller $\tau_f$ in Ref. \cite{Rosti2020}), the centre of mass of the resulting droplet cloud can be expected to shift towards the orifice, in comparison to the no-inertia case.



\begin{figure}[!h]
\centering%
\begin{minipage}{0.49\textwidth}
  \includegraphics[width=1.0\textwidth]{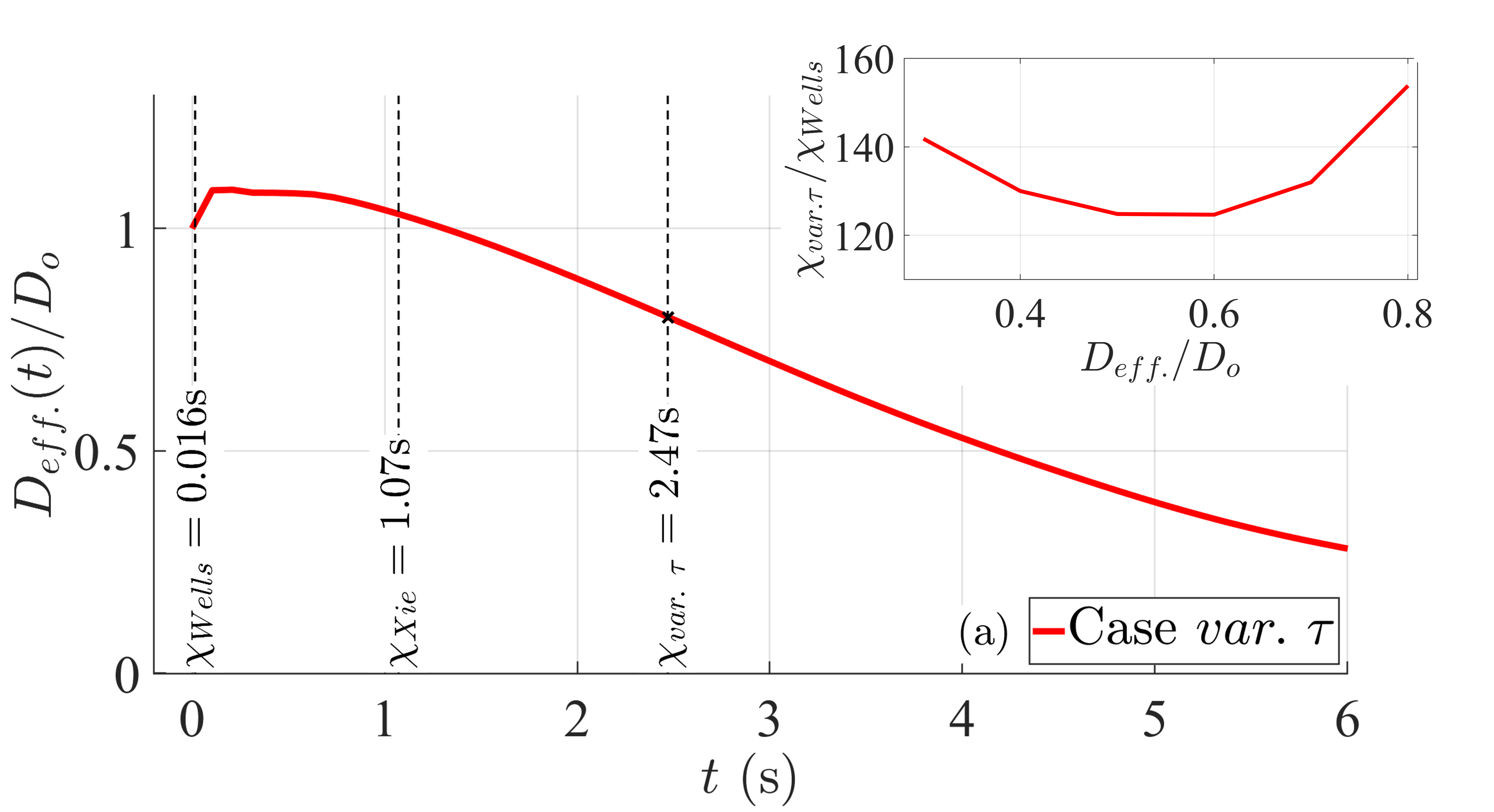}
\end{minipage}%
\hspace{0.01\textwidth}%
\begin{minipage}{0.49\textwidth}
  \includegraphics[width=1.0\textwidth]{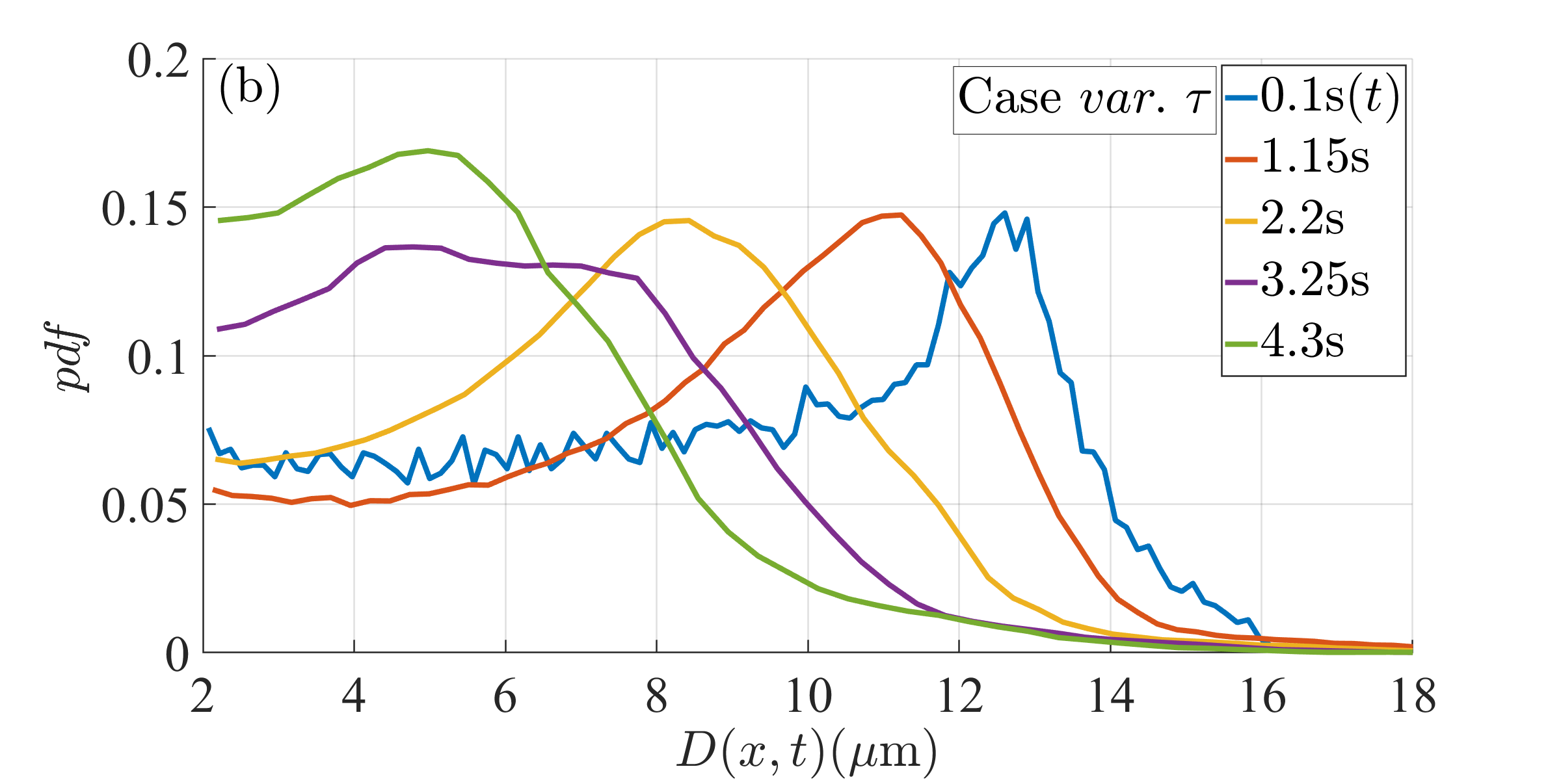}
\end{minipage}
\caption{(a)Time variation of the effective droplet diameter $D_{eff}/D_o$; $D_o$ being the initial diameter. The inset compares the droplet lifetimes in the present simulation with predictions from Wells \cite{WELLS1934} for different shrinkage criteria. (b) Comparison of probability distribution function ($pdf$) for droplet diameters at different times instants. The mean droplet size decreases as the liquid evaporates. }
\label{fig_drop_dia}
\end{figure}

\begin{figure}[!h]
\centering%
\begin{minipage}{0.49\textwidth}
  \includegraphics[width=1.0\textwidth]{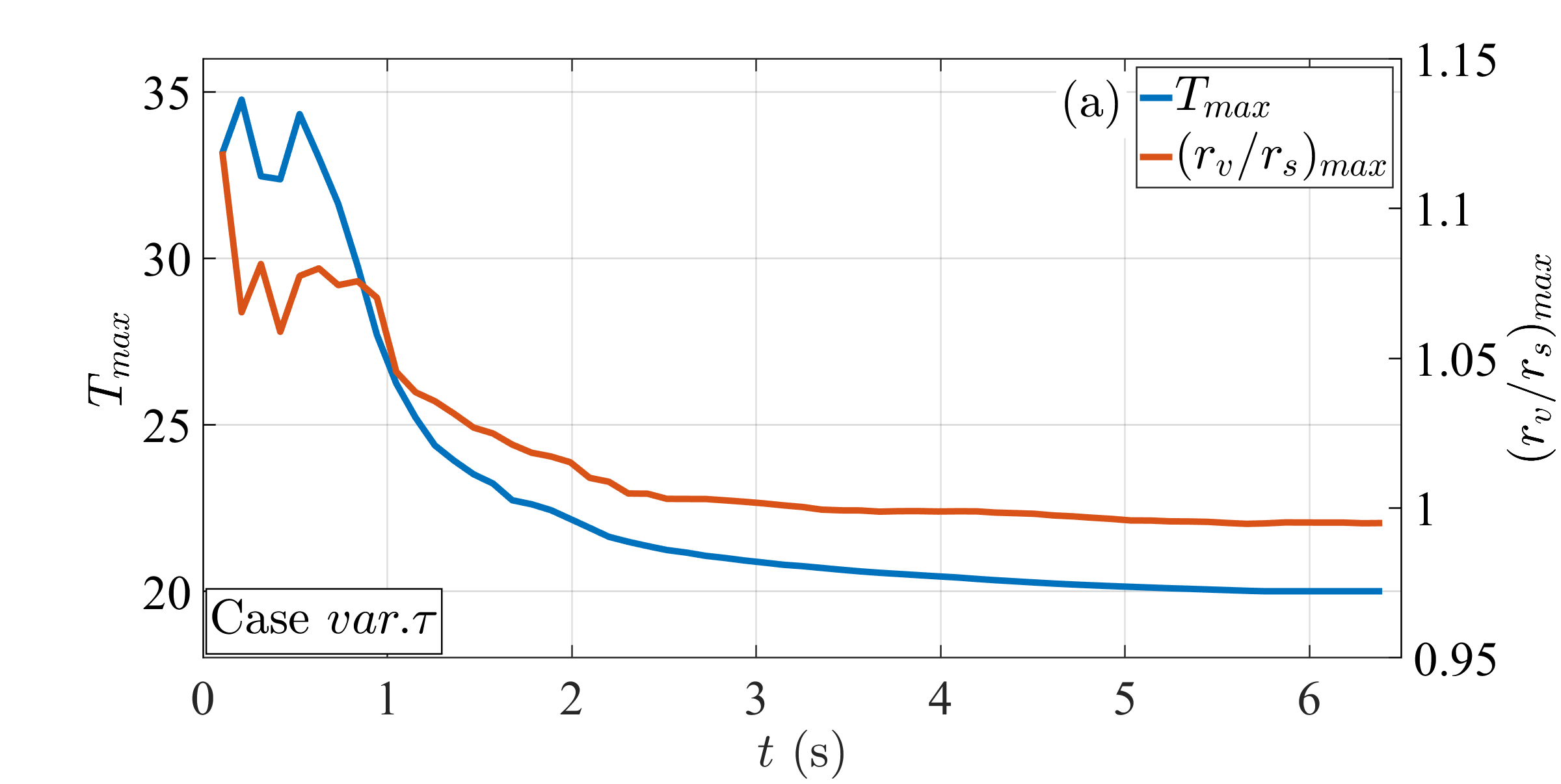}
\end{minipage}%
\hspace{0.01\textwidth}%
\begin{minipage}{0.49\textwidth}
  \includegraphics[width=1.0\textwidth]{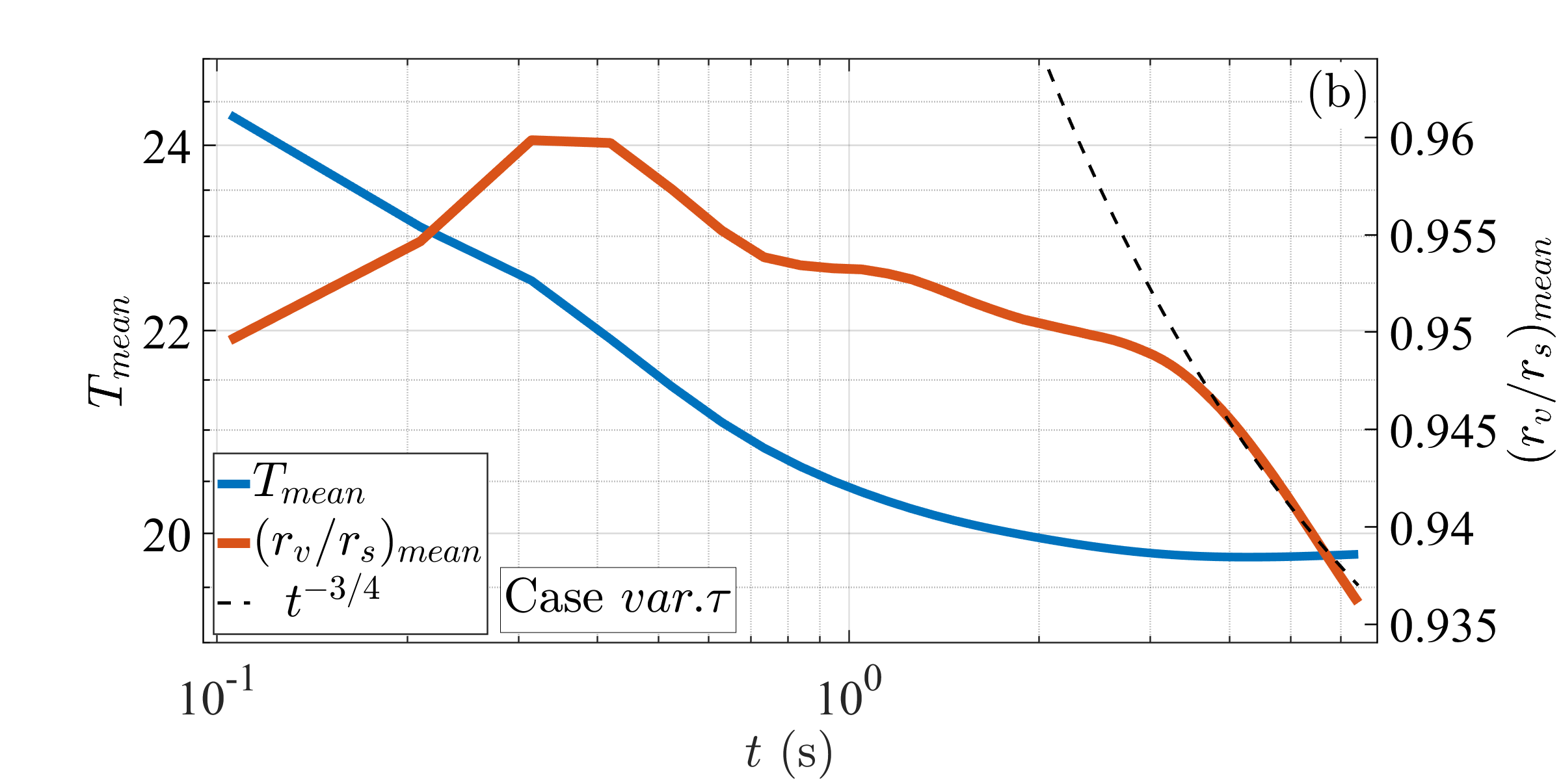}
\end{minipage}
\caption{Time variation of (a) maximum and (b) mean  temperature and relative humidity within the cough volume. }
\label{fig_RH_temp_max_mean}
\end{figure}

\begin{figure}[!h]
\centering%
\begin{minipage}{0.45\textwidth}
  \includegraphics[width=1.0\textwidth]{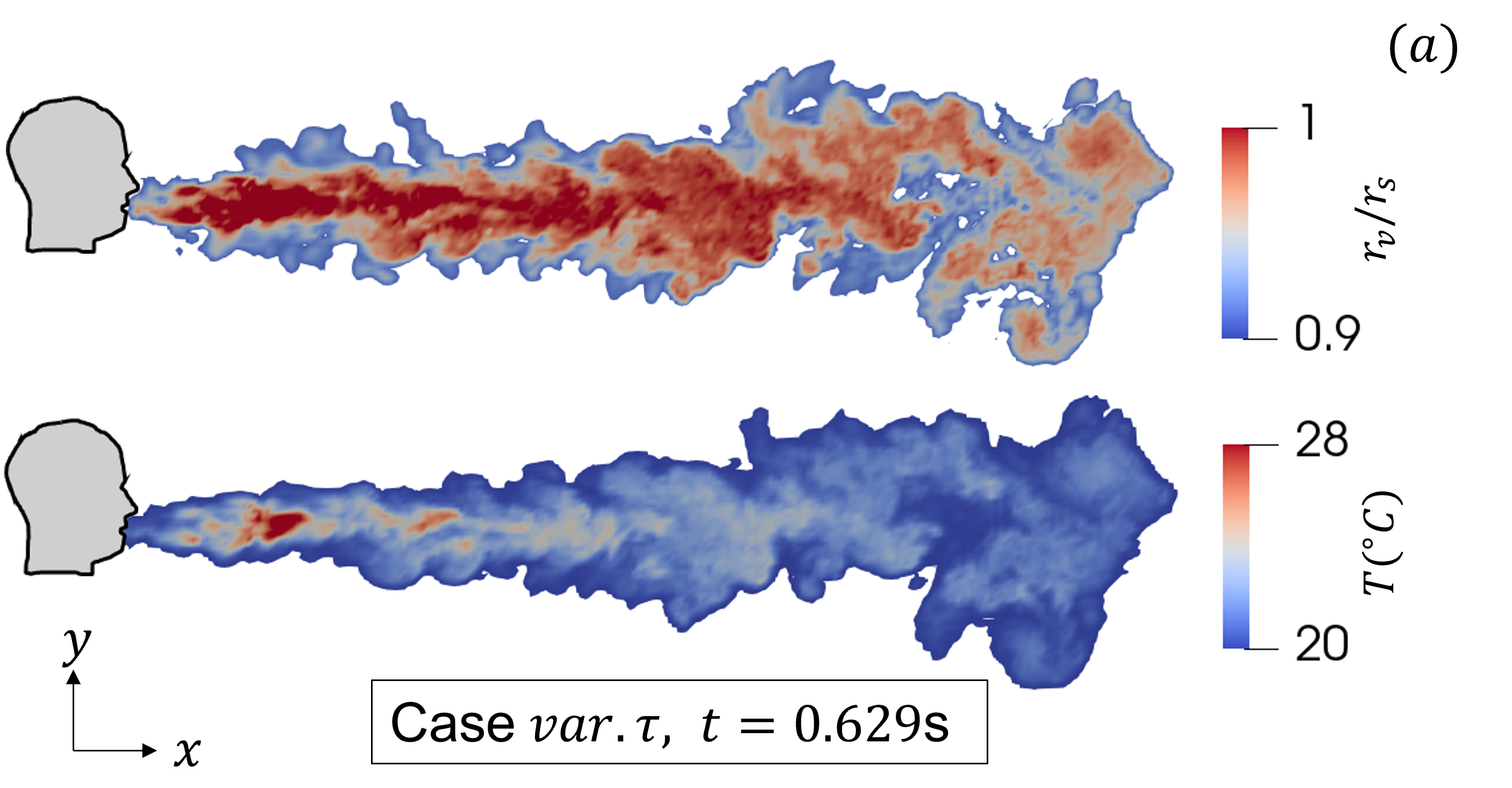}
\end{minipage}%
\hspace{0.01\textwidth}%
\begin{minipage}{0.52\textwidth}
  \includegraphics[width=1.0\textwidth]{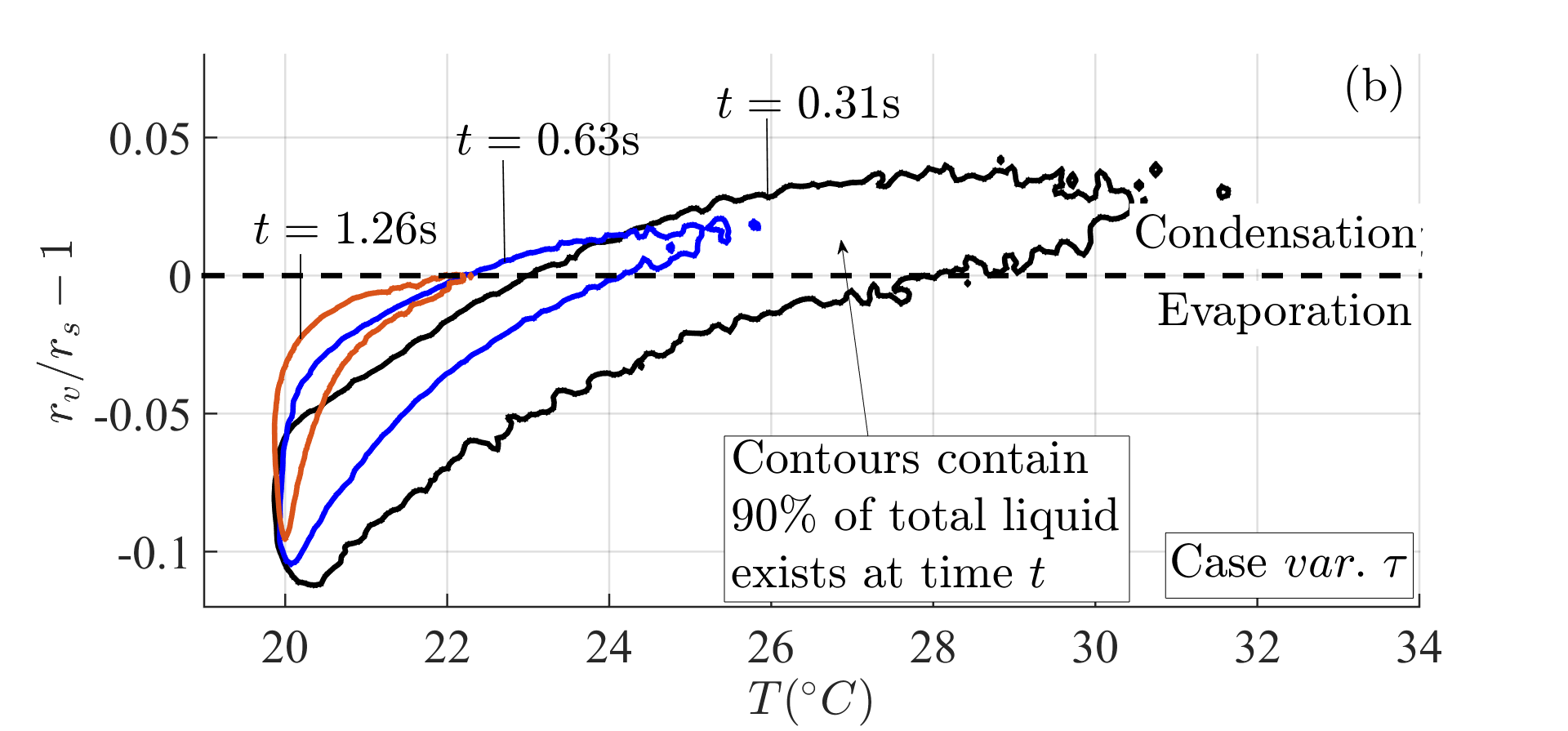}
\end{minipage}
\caption{(a) Contours of relative humidity (top) and temperature (bottom) at time $t=0.629$s. (b)  
Contours wrapping 90\% of total liquid content at time $t$ plotted on a local supersaturation ($r_v/r_s - 1$) vs.  temperature ($T$) diagram. The plot shows the saturation-temperature conditions at which most of the liquid content (or droplets) exists at any time $t$.}
\label{fig_RH_temp}
\end{figure}

Next, we calculate the lifetimes ($\chi$) of droplets from our simulation and compare them with the theoretical estimates of Wells \cite{WELLS1934} and Xie et al. \cite{xie2007far}; also see Ref. \cite{Lohse2021}. Towards this we consider the total liquid content at a given time instant (figure \ref{fig_liq_cont}) to be consisting of $N_o$ droplets of uniform size and calculate the effective droplet diameter $D_{eff}$ by equating the total liquid volume to  $N_o(\pi/6)D^3_{eff}$. The variation of $D_{eff} / D_o$ with time is presented in figure \ref{fig_drop_dia}a. Following Chong et al. \cite{Lohse2021}, we first calculate the $80\%$ lifetime, i.e., the time taken by droplets to shrink to $80\%$ of the initial diameter. This comes out to be $\chi_{var.\tau} = 2.47$s marked in figure \ref{fig_drop_dia}a. Also shown in the figure are the estimates of Wells ($\chi_{Wells}$) and Xie et al. ($\chi_{Xie}$) for the $80\%$ lifetime. We find that $\chi_{var.\tau} / \chi_{Wells} = 153.4$  which is consistent with the range 100-150 obtained for this ratio by Chong et al. \cite{Lohse2021} in their simulations. The accurate capture of extended lifetimes of the cough droplets, vis-a-vis Wells \cite{WELLS1934}, provides a further verification to the present approach. Note that Wells \cite{WELLS1934} derived the dependence of droplet lifetime on its size, considering an isolated droplet evaporating in an ambient whose temperature and relative humidity remain unchanged (see Section S3 in the supplementary material). Since the temperature and relative humidity experienced by droplets within the cough volume can vary considerable in space and time due to turbulence (see figures \ref{fig_RH_temp_max_mean} and \ref{fig_RH_temp} below and also Ref. \cite{Lohse2021}), $\chi_{Wells}$ underestimates the actual droplet lifetime by two orders of magnitude. On the other hand, Xie et al. \cite{xie2007far} coupled the local temperature and humidity fields experienced by a single droplet to a \textit{steady-state} jet in estimating droplet lifetime. As a result $\chi_{Xie}$ is much larger than $\chi_{Wells}$ but still falls short of $\chi_{var.\tau}$ by a factor of 2.4 (figure \ref{fig_drop_dia}a). This points out the need to invest further modelling efforts to predict droplet lifetimes accurately for a realistic cough, which is inherently \textit{transient} in nature.

The 80\% criterion used by Chong et al.\cite{Lohse2021} was possibly due to a short simulation time ($1.2$s) used in their study. In the present work, we have run the simulation till $6$s, which allows us determine the time taken for a droplet to shrink to about 30\% of the initial diameter. We have determined droplet lifetimes for the shrinkage criterion ranging from 80\% to 30\% and compared them with the Wells' estimates, calculated using equation S3 in the supplementary material for these shrinkage criteria. The ratio $\chi_{var.\tau}/\chi_{Wells}$ is plotted as a function of $D_{eff}/D_o$ corresponding to the different shrinkage criteria; see the inset in figure \ref{fig_drop_dia}a. The variation of $\chi_{var.\tau}/\chi_{Wells}$ is non-monotonic, highlighting the crucial role played by the effective temperature and relative-humidity fields surrounding droplets in determining the evaporation rate for a given droplet size. Note that the ratio $\chi_{var.\tau}/\chi_{Wells}$ remains of the order $\sim$100, consistent with previous results.

As discussed in section \ref{subsec:Thermodynamics}, the $var. \tau$ model enables us to calculate the local diameter $D(\bold{x},t)$ based on the local $\tilde{r}_l(\bold{x},t)$ distribution (Eq. \ref{eq:m_li}). Figure \ref{fig_drop_dia}b shows the probability density function ($pdf$) of the droplets (for $D(\bold{x},t)>2\mu$m) at different time instants. A range of droplet sizes up to a maximum of $18\mu$m is present, all of which originate from the mono-disperse initial droplet diameter of $10\mu$m. For $t=0.1$s and $1.15$s, the $pdf$ peaks at diameters greater than $10\mu$m indicating that, for small times, the dominant phase-change process is the condensation of water vapour due supersaturation. This is apparent from figure \ref{fig_RH_temp_max_mean}a, which shows that the maximum relative humidity within the cough flow much exceeds $RH=1$ for $t<2$s. The presence of supersaturation is the primary reason why $D_{eff}/D_o > 1$ for small times (figure \ref{fig_drop_dia}a) and why the droplets have extended lifetimes as seen earlier; see Ref. \cite{Lohse2021}. Referring back to figure \ref{fig_drop_dia}b, for $t \geq 2.2$s, the peak in the $pdf$ shifts to droplets smaller than $10\mu$m suggesting that droplet evaporation becomes the dominant phase-change process, due to an increased sub-saturation of the environment; see figure \ref{fig_RH_temp_max_mean}b.

Figure \ref{fig_RH_temp_max_mean}a shows the maximum temperature ($T_{max}$) and relative humidity ($(r_v/r_s)_{max}$) present within the cough volume; the corresponding mean values ($T_{mean}$, $(r_v/r_s)_{mean}$), averaged over $V(t)$, are plotted in figure \ref{fig_RH_temp_max_mean}b. Due to the presence of large condensation rates during the cough ($t<0.528$s), some regions undergo sufficient heating to raise the maximum temperature slightly above $34^\circ$C, which is the orifice temperature (figure \ref{fig_RH_temp_max_mean}a). After the cough ends, $T_{max}$ declines at a rapid rate primarily due to the entrainment of ambient fluid \cite{diwan2020understanding}; there is also a contribution from the droplet evaporation which acts as a sink for temperature. On the other hand, the maximum and mean relative humidity decreases much more slowly (figures \ref{fig_RH_temp_max_mean}a,b), as the evaporation acts as a source of water vapour. Due to the decrease in temperature (requiring less water vapour for saturation) and continued evaporation, there always exists a region in the flow which is near saturation condition; i.e., $(r_v/r_s)_{max} \approx 1$ even for $t>2.5$s (figure \ref{fig_RH_temp_max_mean}a). On average, however, the cough fluid is sub-saturated for the entire simulation time as seen in figure \ref{fig_RH_temp_max_mean}b. Rosti et al. \cite{Rosti2020} have argued that the mean saturation in a moist puff should decay like $t^{-3/4}$. They observed the $t^{-3/4}$ behaviour for an extended time duration ($0.1-100$s) due to a smaller value (60\%) of the ambient relative humidity used in their simulations, which is not expected to lead to supersaturation \cite{Lohse2021}. In the present case, the $t^{-3/4}$ variation for $(r_v/r_s)_{mean}$ is apparent after $t=3$s (figure \ref{fig_RH_temp_max_mean}b), presumably because of the presence of supersaturation in some parts of the flow at earlier time instants. Interestingly, the start of this power-law decay coincides with the location where $T_{mean}$ nearly reaches a constant value close to the ambient temperature. One may expect that the $t^{-3/4}$ variation would continue if the present simulation was run for a longer time. A similar behaviour of $(r_v/r_s)_{mean}$ is also observed for the $const. \tau$ case, as shown in figure S3 in the supplementary material.

In the following, we present detailed distributions of the various flow and thermodynamic quantities, which is the real strength of the present computational scheme. Figure \ref{fig_RH_temp}a shows contour plots of the relative humidity and temperature at $t=0.629$s, i.e., just after the cough ends; both exhibit a considerable degree of inhomogeneity due to turbulence. At this time instant, the region within the core of the cough flow is supersaturated (indicated by saturated red colour in the colour map; figure \ref{fig_RH_temp}a) whereas the region near the edges of the flow is in a sub-saturated state; see also Ref. \cite{Lohse2021}. Thus both condensation and evaporation take place simultaneously within the cough volume. This is seen more clearly in figure \ref{fig_RH_temp}b which presents the evolution of the phase-change characteristics on a supersaturation ($r_v/r_s -1$) vs. temperature plot. This plot is obtained as follows. The total liquid content present within the cough volume ($\tilde{r}_v>0.91$) at a given time instant is distributed over 250$\times$250 bins of the saturation-temperature grid and is summed over each bin. A specific contour level is chosen such that 90\% of the total liquid present at that time instant lies inside that contour. Such contours are plotted in figure \ref{fig_RH_temp}b for three time instants. For example, at time $t=0.31$s, 90\% of the liquid content (or droplets) experience the supersaturation-temperature conditions that lie inside the black curve in figure \ref{fig_RH_temp}b. The dashed line at $r_v/r_s =1$ (figure \ref{fig_RH_temp}b) provides a demarcation between condensation (due to supersaturation) and evaporation (due to sub-saturation). These can be mapped onto the regions in physical space affected by condensation/evaporation by making contour plots such as the one presented in figure \ref{fig_RH_temp}a.  At $t=0.31$s in figure \ref{fig_RH_temp}b, a considerable region is subjected to supersaturation causing an increase in liquid content, consistent with the previous discussion in relation to figures \ref{fig_drop_dia} and \ref{fig_RH_temp_max_mean}. As time increases, the contour contracts more rapidly along the temperature axis as compared to the saturation axis. Furthermore, the effect of supersaturation (i.e., $r_v/r_s > 1$) diminishes rapidly with time so that beyond at $t=1.26$s, the liquid content is expected to decay by evaporation; see figure \ref{fig_liq_cont}. The saturation-temperature diagram in figure \ref{fig_RH_temp}b can serve as a useful tool in visualizing the thermodynamics of phase change in an evolving cough flow.

\begin{figure}[!h]
\centering%
\begin{minipage}{0.46\textwidth}
  \includegraphics[width=1.0\textwidth]{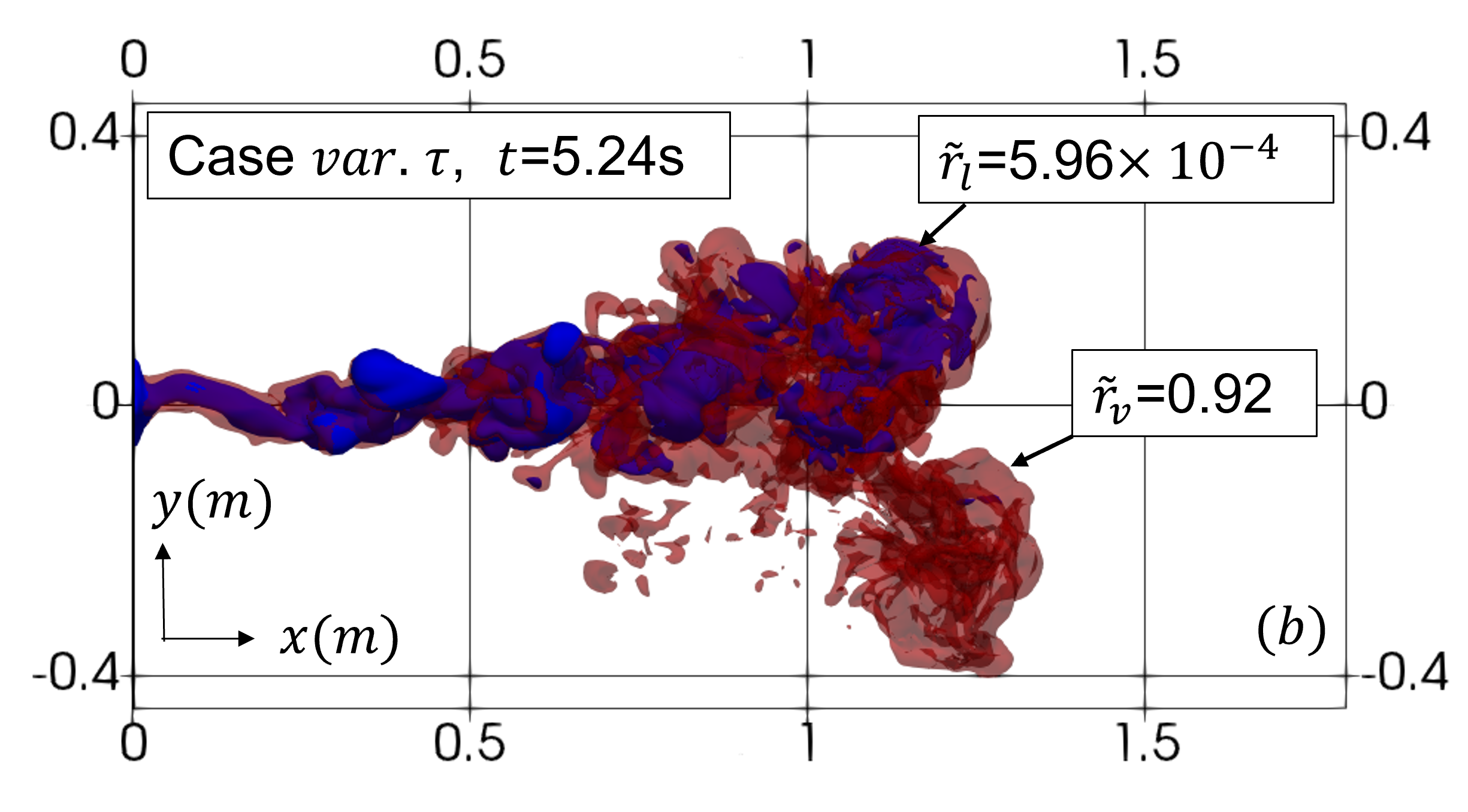}
\end{minipage}
\hspace{0.00\textwidth}%
\begin{minipage}{0.52\textwidth}
  \includegraphics[width=1.0\textwidth]{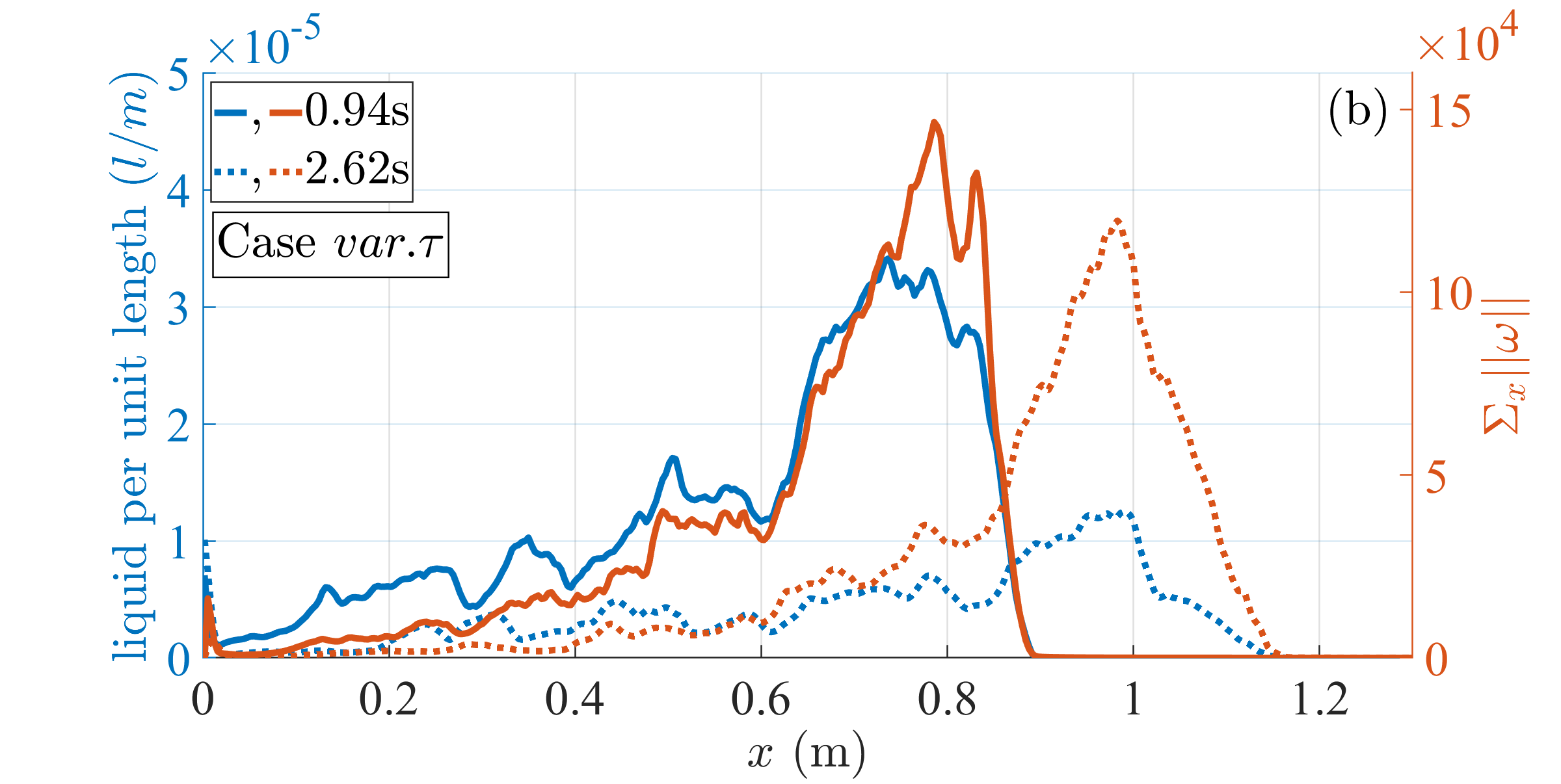}
\end{minipage}
\caption{(a) Iso-surfaces of the vapour mixing ratio at $\tilde{r}_v=0.92$ (semi-transparent red) and the liquid mixing ratio (blue) at $\tilde{r}_l=5.96\times 10^{-4}$. (b) The axial distribution of liquid content per unit length and total vorticity at a given $x$ location (summed over the $y-z$ plane). These plots for the $var.\tau$ model can be compared with those for the $const.\tau$ model presented in figure S4 of the supplementary material. In particular, figure S4(a) shows that the chunk separated from the main flow for the $const.\tau$ model moves to greater streamwise distances than that seen here for the $var.\tau$ model.
}
\label{fig_axial_liquid}
\end{figure}

An iso-surface of water vapour with $\tilde{r}_v=0.92$ at $t=5.24$s is shown in figure \ref{fig_axial_liquid}a. The slightly higher value of $\tilde{r}_v=0.92$ chosen here (as compared to 0.91 used earlier as a threshold for determining $V(t)$) enables discerning certain flow features better, especially the chunk of fluid which is about to separate from the main cough flow at this time instant seen in figure  \ref{fig_axial_liquid}a. This feature has been reported by Liu et al. \cite{balachandar2021investigation}, who found that the separated portion of the cough flow carries droplets with it and travels at a faster speed than the main flow, potentially resulting in a faster spread of infection. Figure \ref{fig_axial_liquid}a also shows an iso-surface of liquid water (blue) with $\tilde{r}_l=5.96\times 10^{-4}$, which is visualized through the semi-transparent vapour contour (red). The vapour is seen to wrap around the liquid content and shield it from the ambient conditions, providing a visual confirmation to the inference drawn in Chong et al. \cite{Lohse2021} that this vapour shielding effect contributes to the extended droplet lifetimes by providing an elevated local relative-humidity field. Note that the liquid mixing ratio of $\tilde{r}_l=5.96\times 10^{-4}$ corresponds to the droplet diameter of $4\mu$m (Eq. \ref{eq:m_li}); droplets of larger size are expected to be present within the central region of the cough flow.

Figure \ref{fig_axial_liquid}b presents the liquid content per unit length and the absolute vorticity magnitude (summed over the $y-z$ plane) as a function of $x$ for two time instants. It is seen that the majority of liquid is concentrated within the ``head'' of the cough flow, which is also the region where vorticity magnitude is large. In fact the streamwise distributions of the liquid concentration and vorticity are similar to each other (figure \ref{fig_axial_liquid}b). The large vorticity present within the cough head suggests presence of a torroidal vortex which can trap liquid droplets within it. Since such vortex structures are known to  maintain their identity over long distances, we may expect the trapped liquid droplets to travel with them and contribute to the long-range transmission of pathogens; see also Ref. \cite{balachandar2021investigation}. The interaction between the liquid content and vorticity fields represents an important aspect of the dynamics of a moist puff, and is being investigated further.

\subsection{Advantages and limitations of the present approach}

The coarse-graining of liquid droplets into an Eulerian liquid field is conceptually simpler and computationally easier to implement as compared to the Eularian-Lagrangian solvers which track individual droplets \cite{Rosti2020,LohsePRF2021growth}. For example, the Eulerian field approximation supports specification of any profile of liquid content at the orifice as a function of $y$, $z$ and $t$. Especially, when the number of droplets in a cough become large, Lagrangian tracking can become computationally expensive. Note that the number of droplets used in the simulations of Rosti et al. \cite{Rosti2020} and Ng et al. \cite{LohsePRF2021growth} is 5000 (in the size range $10-1000 \mu$m) whereas the equivalent number of droplets (of diameter $10\mu$m) in the present simulation is $\sim 2 \times 10^7$. The latter corresponds to a droplet concentration of $\sim 3 \times 10^4 \ \mathrm{cm^{-1}}$ (in terms of the expelled cough volume), which is relevant for realistic coughs with maximum droplet diameters of $10-20 \mu$m; see Ref. \cite{bourouiba2021review}. The present approach is particularly suitable for such situations. Even for coughs wherein a wide range of initial droplet sizes is present, the flow trajectory is not much affected by the gravitational settling of larger droplets and the droplets which continue to evolve with the flow are typically less than $50 \mu$m \cite{Lohse2021}. As such droplets move further from the orifice, the Stokes number associated with them reduces to sufficiently low values (figure \ref{fig_st}), due to increased local flow time scale, so that the droplets are effectively carried by the flow. Our computational scheme can accurately capture this regime, which is relevant for determining the long-range virus transmission. The total cough liquid (laden with virions) ingested by a susceptible person, standing a few feet away from an infected person, determines the probability of infection, which is an important input to epidemiological modelling \cite{domino2021case, abinito_model}. This is another area where simulations of the kind presented here will of value.

The limitations of the present approach should also be noted. Firstly, the use of the modified one-moment scheme implemented here requires a closure assumption relating the local droplet number density and the local droplet size. Our choice for this closure for the $var.\tau$ model,  Eq. (\ref{eq:tau_ci}), is to assume that the number density of droplets is uniform in the cough volume. Secondly, our approach works for a mono-disperse distribution of droplets released from the orifice, whereas respiratory events are known to produce a wide range of droplet sizes as discussed above. Thirdly, we do not incorporate droplet inertia and settling effects and therefore preclude large-sized droplets, with higher Stokes numbers, from our analysis. Finally, we implicitly assume that there are no aerosol particles (or condensation nuclei) in the ambient air, which may not always be true. 

Some of these limitations can be overcome, to varying degrees, by introducing additional considerations. A scalar transport equation may be solved for (an) additional moment(s) of the droplet size distribution to obtain its distribution in space and time \cite{beck2002development}. The droplet inertia and gravitational settling can be introduced (within the liquid field approximation) for larger droplets with small but finite Stokes numbers using Eq. (\ref{eq:maxey_st}) above. Moreover, additional liquid water scalars representing different initial droplet sizes could be used, in analogy with multiple hydrometeor classes in simulations of atmospheric flows (e.g. \cite{hernandez2013minimal}), although at the cost of simplicity. A development of the present solver along some the lines mentioned here is being carried out.

\section{Conclusion} \label{sec:conclusion}
We have used a computational approach (that may be called the extended one-moment scheme) for simulating moist cough flows, that uses an Eulerian field approximation for the liquid droplets, which are advected by fluid streamlines. A closure model is proposed for the time scale of evaporation ($\tilde{\tau}_c$), which is a crucial parameter for the thermodynamics of phase change. The model considers the droplet density to be uniform in space (but varying in time) and the droplet radius a function of space and time. This model (called the $var. \tau$ model) is compared with the $const. \tau$ model, in which $\tilde{\tau}_c$ is kept constant. We find that the liquid content within the cough flow takes much longer to evaporate for the $var. \tau$ model, as compared to that for the $const. \tau$ model, due to a gradual increase in  $\tilde{\tau}_c$ with time for the former, caused by dilution and evaporation. The $var. \tau$ model is shown to be closer to the realistic cough flow scenarios and has been used for a detailed investigation in the context of a mild cough. The Stokes numbers associated with the initial droplet diameter of $10\mu$m are plotted as a function of space and time, for three different measures of the flow time scale. The Stokes number based on the Kolmogorov time scale is found to be of the order of $10^{-2}$ or smaller, providing a verification to our assumption that the liquid-droplet inertia is negligible and that the liquid field is carried by the local fluid velocity. A careful analysis of the evolution and thermodynamics of the cough flow has been carried out, and a comparison is made with the results available in the literature. The following aspects of the available results have been successfully reproduced.
\begin{itemize}
    \item The mean velocity and the streamwise extent of the cough flow follow the $t^{-3/4}$ and $t^{1/4}$ variations, respectively. The mean relative humidity follows the $t^{-3/4}$ law for the latter part of the flow evolution \cite{Rosti2020}.
    \item The liquid field in the core of the cough flow undergoes an initial condensation followed by evaporation at later times \cite{Lohse2021}. This is due to the initial supersaturation on account of a high ambient relative humidity ($90\%$).
    \item The effective droplet diameter shows an extended lifetime, which is 120-150 times larger than that predicted by the Wells' model \cite{WELLS1934}; see Ref. \cite{Lohse2021}.
    \item At later time instants in the flow evolution, a chunk of fluid separates from the main cough flow and causes an increased spread of the flow in the lateral/streamwise direction \cite{balachandar2021investigation}.
\end{itemize}

This comparison provides a strong support to the utility of our proposed approach, and we have used it to obtain some new results. For the ``mild'' cough simulated herein (with the initial droplet diameter of $10\mu$m) the time taken for the liquid content to reach to 5\% of its initial value has been found to be about 7-10 times the cough duration. The interpretation of the liquid field in terms of local droplet sizes (for the $var.\tau$ model) has enabled us to plot a probability density function for droplets sizes and track its evolution with time. A saturation-temperature diagram has been constructed at different time instants, which shows that the temperature of the cough flow decreases with time more rapidly than the relative humidity, thereby promoting conditions of supersaturation close to the orifice. Finally, a portion of the liquid content is shown to be trapped within the head of the cough flow, which is also a region of high vorticity, and is therefore likely to survive over longer distances from the orifice. We believe the present approach is well suited to study the long-range transport of small droplets in a cough flow, which are responsible for the airborne transmission of the COVID-19 type pathogens. Our results can also provide useful inputs for calculating infection probabilities \cite{singhal2021virus}, in the context of epidemiological modelling.

\paragraph{Acknowledgements}
The present simulations have been carried out at the Supercomputer Education and Research Centre at Indian Institute of Science, Bengaluru. SR is supported through the Swedish Research Council grant no. 638-2013-9243. SSD acknowledges support from Indian Institute of Science, Bengaluru towards running the simulations. The authors declare no competing interests.

\paragraph{Data Availability Statement} The data that support the findings of this study are available on request from the corresponding author, SSD.

\paragraph{Supplementary Material} The present paper contains online supplementary material.

\bibliographystyle{unsrt} 
\bibliography{references}
\end{document}